\documentclass[12pt]{article}
\usepackage[margin=0.75in]{geometry}
\usepackage{authblk}

\usepackage{amsfonts}
\usepackage{amsmath}
\usepackage{amssymb}
\usepackage{physics}
\usepackage{braket}
\usepackage{slashed}
\usepackage{mathtools}
\usepackage{url}
\usepackage{breqn}
\usepackage{color}
\usepackage{array}
\usepackage{graphicx}
\usepackage{dsfont}
\usepackage{bbm}
\usepackage{subcaption}
\usepackage{float}
\usepackage{bm}
\usepackage{upgreek}
\usepackage[labelfont=bf, font = small]{caption}

\usepackage[linktocpage]{hyperref}
\hypersetup{
    colorlinks = true,
    urlcolor = magenta,
    linkcolor = blue,
    citecolor = magenta,
    }

\usepackage[sorting=none, style=numeric-comp,giveninits,maxbibnames=4,maxcitenames=4]{biblatex}
\renewbibmacro{in:}{}
\DeclareFieldFormat{pages}{#1}
\DeclareFieldFormat[article]{title}{}
\DeclareFieldFormat[article]{journaltitle}{#1}
\DeclareFieldFormat[inproceedings]{title}{}
\DeclareFieldFormat[inproceedings]{booktitle}{In \textit{#1}}
\DeclareFieldFormat[inproceedings]{pages}{pp. #1}

\title{\textbf{Non-thermal WIMP Production from Higher Order Moduli Decay}}
\author[1,2]{Amitayus Banik\thanks{E-mail: \href{mailto:abanik@cbnu.ac.kr}{abanik@cbnu.ac.kr}}}
\author[3]{Manuel Drees\thanks{E-mail: \href{mailto:drees@th.physik.uni-bonn.de}{drees@th.physik.uni-bonn.de}}}
\affil[1]{\small \textit{Institut f\"{u}r Theoretische Physik und Astrophysik,
    Universit\"{a}t W\"{u}rzburg, Campus Hubland Nord, Emil-Hilb-Weg 22,
    D-97074 W\"{u}rzburg, Germany}\vspace{2mm}}
\affil[2]{\small \textit{Department of Physics, Chungbuk National University, Cheongju, Chungbuk 28644, Korea}\vspace{2mm}}
\affil[3]{\small \textit{Bethe Center for Theoretical Physics and Physikalisches
    Institut, Universit\"{a}t Bonn, Nussallee 12, D-53115 Bonn, Germany}}

\date{}

\begin{document}
\def\lsim{\:\raisebox{-0.5ex}{$\stackrel{\textstyle<}{\sim}$}\:}
\def\gsim{\:\raisebox{-0.5ex}{$\stackrel{\textstyle>}{\sim}$}\:}

\maketitle

\begin{abstract}
  \noindent In a non--standard cosmological scenario heavy,
  long--lived particles, which we call moduli, dominate the energy
  density prior to Big Bang Nucleosynthesis. Weakly Interacting
  Massive Particles (WIMPs) may be produced non-thermally from moduli
  decays. The final relic abundance then depends on additional
  parameters such as the branching ratio of moduli to WIMPs and the
  modulus mass. This is of interest for WIMP candidates, such as a
  bino--like neutralino, where thermal production in standard
  cosmology leads to an overdensity. Previous works have shown that
  the correct dark matter (DM) relic density can then still be
  obtained if the moduli, with mass less than $10^{7} \,\text{GeV}$,
  decay to WIMPs with a branching ratio of less than $10^{-4}$. This
  upper bound could easily be violated once higher order corrections,
  involving final states with more than two particles, are included.
  We compute the branching ratios of three-- and four--body decays of
  a modulus into final states involving two DM particles for general
  couplings. We then apply these expressions to sparticle production
  within the Minimal Supersymmetric Standard Model (MSSM) with
  neutralino DM. We find that this upper bound on the branching ratio
  can be satisfied in simplified models through an appropriate choice
  of as yet undetermined couplings. However, in the MSSM, it requires
  sparticle masses to be very close to half the modulus mass, in
  contrast to the idea of weak--scale supersymmetry.
\end{abstract}

\newpage
\hrule
\tableofcontents 
\vspace{5mm}
\hrule

\section{Introduction}

Although the existence of Dark Matter (DM) has been consensus among
cosmologists for several decades and is a crucial ingredient of the
current standard model of cosmology, its precise nature remains
unknown. Proposals for its explanation generally require the advent of
new physics \cite{Bertone:2004pz}, as the Standard Model (SM) of
particle physics does not contain a suitable candidate. Many
extensions of the SM that can accommodate viable DM candidates have
been suggested.

In particular, Weakly Interacting Massive Particles (WIMPs) have been
extensively studied as DM candidates. They can be produced thermally:
in standard cosmology their decoupling, or freeze--out, from the
thermal plasma of SM particles due to the expansion of the universe,
leads to the correct present-day relic for roughly weak (i.e.
${\cal O}(\rm pb)$) annihilation cross sections into (some) SM
particles. WIMPs can thus be searched for in both direct and indirect
detection experiments \cite{PDG}. However, no conclusive evidence has
been found to date for their existence, which has led to interest in
exploring scenarios beyond the usual thermal WIMP.

One such scenario involves a modified cosmological history with an era
of early matter domination. This is motivated by UV complete theories.
For example, supersymmetry breaking in supergravity (SUGRA) theories
usually involve a hidden sector \cite{Bailin:2004zd}. It contains
``Polonyi fields'' which couple very weakly to SM particles and are
thus long lived. Moreover, superstring theory typically contains many
very weakly coupled scalar ``moduli'' fields arising from, for
example, the compactification of extra dimensions
\cite{Polchinski:1998rq, Polchinski:1998rr}. For our purposes, the most
relevant such field is the one with the longest lifetime; since the
lifetime scales like an inverse power of the mass, this will often be
the lightest scalar particle in this sector, which we call the modulus for
definiteness.

Such fields can acquire large values during inflation
\cite{Dine:1995uk}, if their mass was smaller than the Hubble
parameter. Once the post--inflationary Hubble parameter becomes
smaller than the modulus mass, this field begins to oscillate
coherently about its minimum; this is equivalent to the modulus field
``condensing'' into an ensemble of non--relativistic particles. We are interested in
this stage of the evolution of the universe, i.e. our starting point is an
ensemble of very massive and long--lived particles. 
Due to their long lifetime this can lead to a period of early matter
domination \footnote{At this point,
possible self--interactions of the massive particle are no longer
relevant; only the mass and lifetime matter, once we have assumed that
these particles dominate the energy content of the universe at some
stage.}. As the universe keeps expanding the matter energy density
decreases more slowly than the radiation energy density; hence this
matter dominated era has to end via decays of the modulus, otherwise
they would have continued to dominate the universe till the present
day, contrary to observation. These decays inject very energetic
particles; this would spoil the largely successful predictions of Big
Bang Nucleosynthesis (BBN) unless this epoch ends sufficiently early,
with a reheat temperature $(T_{\text{RH}})$ of at least 4 MeV
\cite{deSalas:2015glj}.

Such an early matter dominated epoch can greatly affect the WIMP relic
abundance. Modulus particles decay out of equilibrium, leading to an
increase of the co-moving entropy density and hence to the dilution of
any pre--existing WIMP density. On the other hand, direct moduli decay
to WIMPs can instead increase the WIMP density. The final WIMP relic
abundance will therefore depend on the mass, lifetime and branching
ratios of the modulus. Thus, the WIMP relic density may be larger or
smaller than in standard cosmology. In fact, if these parameters can
be chosen freely, any WIMP model can be made ``safe''
\cite{Gelmini:2006pw}.

The relic abundance of WIMPs in the case of an early matter dominated
era has been widely studied \cite{Chung:1998rq, Giudice:2000ex,
  Allahverdi:2002nb, Pallis:2004yy, Kane:2015qea, Allahverdi:2018aux}. In addition to
dilution by entropy injection and direct WIMP production from modulus
decays, one in general also has to consider WIMP production from the
(subdominant) thermal plasma at temperature $T > T_{\rm RH}$. Consider
a DM WIMP whose thermal relic density is too high. An epoch of early
matter domination can reduce this prediction to an acceptable level
only if two conditions are satisfied.  First, $T_{\rm RH}$ needs to be
below the freeze--out temperature $T_F$, typically
$T_{\rm RH} < m_{\rm DM}/25$, otherwise the WIMP would thermalize in
the radiation--dominated era after modulus decay, and the prediction
of standard cosmology would be recovered. Secondly, the contribution
from direct modulus decays must be sufficiently small
\cite{Drees:2017iod}.  It is given by
\begin{equation} \label{dmdensity}
  \Omega_{\text{DM}}^{\Phi \ {\rm dec}} h^2 \simeq 0.12 \frac{T_{\text{RH}}}
  {1 \ {\rm GeV}} \frac {B_{\Phi \rightarrow \text{DM}}} {10^{-5}}
  \frac{ 10^6 \ {\rm GeV}} {m_\Phi} \frac {m_{\rm DM}} {30 \ {\rm GeV}}\,.
\end{equation}
Here $m_\Phi$ is the mass of the modulus $\Phi$,
$B_{\Phi \rightarrow \text{DM}}$ is the effective branching ratio of
moduli decay to WIMPs (i.e. the average number of WIMPs produced in
the decay of a $\Phi$ particle), and $h \simeq 0.7$ is the Hubble
constant in units of $100 \ {\rm km}/({\rm s}\cdot {\rm Mpc})$. The
dependence on $m_\Phi$ and $T_{\rm RH}$ can be understood from the
consideration that the $\Phi$ number density just before its decay at
$T = T_{\rm RH}$ scales like $T_{\rm RH}^4 / m_\Phi$ while the entropy
density (to which the WIMP density is normalized) scales like
$T_{\rm RH}^3$. The reheat temperature can be computed uniquely from
the total modulus decay width,
$T_{\rm RH} \propto \sqrt{\Gamma_\Phi M_{\rm Pl}}$. If the modulus
decays via a dimension 5 interaction suppressed by the Planck mass
$M_{\rm Pl} = 1.22\cdot 10^{19}$ GeV, then
\begin{equation} \label{phiwidth}
  \Gamma_\Phi = C \frac {m_\Phi^3} { 8 \pi M_{\rm Pl}^2}\,,
\end{equation}
where $C$ is a dimensionless number which depends on the square of the
coefficient of the dimension--5 operator as well as on the number of
different final states (e.g. color factors). The numerical analysis of
ref.\cite{Drees:2017iod} concluded that for $C = 1$,
\begin{equation} \label{bbound}
  B_{\Phi \rightarrow \text{DM}} \lsim 10^{-4} \frac  {100 \ {\rm GeV}} {m_{\rm DM}}
\end{equation}
is required. However, eq.(\ref{dmdensity}) shows that quite generally
$B_{\Phi \rightarrow \text{DM}} \ll 1$ is required for
$m_\Phi \leq 10^7$ GeV, since BBN implies $T_{\rm RH} > 4$ MeV, as
noted above.

An immediate consequence is that the main $\Phi$ decay mode(s) should
{\em not} involve DM particles. However, as pointed out in
ref.\cite{Allahverdi:2002nb}, DM particles may then still be produced
in higher order diagrams, e.g. involving the exchange of off--shell
mediators between the dark sector and SM particles. In this work, we
explicitly evaluate the branching ratios for such higher--order decays
involving final states with three or four particles. We find that
the branching ratio can be sufficiently small if either the relevant
couplings are well below unity, or if the DM mass is very close to
$m_\Phi/2$. The former can be arranged in models introducing new
mediator particles with essentially arbitrary couplings. In contrast,
in supersymmetric models {\em all} superparticles will rather quickly
decay into DM particles. Since they can be produced via ${\cal O}(1)$
couplings the branching ratio can be sufficiently suppressed only if
sparticle masses are very close to, or above, $m_\Phi/2$.

In our explicit calculation, we assume the ``modulus'' particle to be
a spinless particle which is a singlet under the SM gauge group, as for
genuine stringy moduli or Polonyi--like particles. We only allow couplings
that respect the full gauge symmetry of the SM; for example, renormalizable
couplings to SM $f \bar f$ pairs are not possible. In the supersymmetric
context, we focus on modulus couplings that arise as $F-$terms. The bound
(\ref{bbound}) can then only be satisfied if the couplings are constructed
such that decays into superparticles are forbidden, or at least strongly
suppressed, in leading order. As already mentioned, higher order diagrams
will then generically still violate this bound.

The rest of this article is organized as follows: in Section
\ref{moddecays}, we compute the branching ratios for WIMP production
via moduli decays in three-- and four--body final states. We work with
simplified models and provide explicit expressions for all relevant
partial widths; we also provide analytical approximations for the
integrated partial widths in the limit $m_{\rm DM}^2 \ll m_\Phi^2$.
In Section \ref{moddecayssusy}, we examine the case of the MSSM with a
gauge singlet modulus superfield and discuss couplings to the particle
spectrum. Taking our WIMP candidate to be the lightest neutralino, we
study sparticle production through moduli decay and ascertain the
parameter space that can satisfy the upper bound on the branching
ratio. We conclude in Section \ref{conclude}. Finally, several
appendices contain details of the computations.


\section{Moduli Decays} \label{moddecays}

We introduce a real massive modulus $\Phi$, which is a singlet under
the SM gauge group. Possible gauge invariant couplings include:
\begin{equation} \label{modSM}
  \mathcal{L}_{\Phi-\text{SM}} = -\frac{C_g}{4\Lambda} \Phi F_{\mu\nu}F^{\mu\nu}
  + C_S \Phi |S|^2
  + \frac{C_f}{\Lambda}\Phi \left[ (S \cdot \overline{f_L})f_R + \text{h.c.}
  \right] \,.
\end{equation}
Here $F_{\mu \nu}$ is the field strength tensor for any gauge field
$A^{\mu}$, which may or may not be part of the SM; $S$ is a scalar
$SU(2)$ doublet like the SM Higgs boson; and $f$ is any SM
fermion. For each possible two--body final state we have written the
operator of lowest possible dimension, with $\Lambda$ as the cut--off
scale for the non--renormalizable operators; this is generally
$\mathcal{O}(\overline M_{\text{Pl}})$ in SUGRA theories, where
$\overline M_{\rm Pl} = M_{\rm Pl} / \sqrt{8\pi} \simeq 2.4 \cdot
10^{18}$ GeV is the reduced Planck mass. In this section, we will
assume that $S$ is the single Higgs doublet of the SM. In unitary
gauge the Lagrangian (\ref{modSM}) then becomes:
\begin{equation}   \label{unimodSM}
  \mathcal{L}_{\Phi-\text{SM}} = -\frac{C_g}{4\Lambda}\Phi F_{\mu\nu}F^{\mu\nu}
  + \frac{C_S}{2} \Phi s^2
  + \frac{C_f}{\sqrt{2}\Lambda}\Phi (s \bar{f}f)
  + \frac{C_f v_s}{\sqrt{2}\Lambda}\Phi \bar{f}f \;,
\end{equation}
where $s$ denotes the real, physical scalar Higgs field, and
$v_s \simeq 246$ GeV is the vacuum expectation value (VEV) associated
with the breakdown of $SU(2) \times U(1)_Y$. Note that the coupling
$C_f$ leads to both two-- and three--body final states already at
leading order. On the other hand, $C_S$ multiplies a dimension--three
operator, and thus has mass units. If $C_S = {\cal O}(m_\Phi)$ the
width for $\Phi \rightarrow s s$ decay is very large, i.e. $\Phi$
decays much too rapidly to ever dominate the energy density of the
universe. We will instead implicitly assume that
$C_S \lsim m_\Phi^2 / \Lambda$; all three terms in the Lagrangian (\ref{modSM})
will then lead to similar $\Phi$ lifetime, if $C_g, \, C_f \lsim 1$.

If $\Lambda$ is near the Planck scale, $\Phi$ decay will be completed
before BBN only if $m_\Phi > 10$ TeV, which allows us to neglect the
masses of SM particles in the kinematics. Hence, using the interaction
vertices of (\ref{modSM}), we may write down the leading order decay
rates of the modulus to SM particles. We begin with the decay into scalars:
\begin{equation} \label{scalardecay}
 \Gamma_{\Phi \rightarrow ss} = \frac{C^2_S}{32\pi m_{\Phi}} \,.
\end{equation}
The corresponding expression for decays into gauge bosons is:
\begin{equation} \label{gaugedecay}
 \Gamma_{\Phi \rightarrow A A} = \frac{N_g C^2_g\,m^3_{\Phi}}{64\pi \Lambda^2} \,,
\end{equation}
where $N_g$ is the dimensionality of the adjoint representation of the
gauge group in question, e.g. $N_g = N^2 - 1$ for an $SU(N)$ gauge
group while $N_g = 1$ for a $U(1)$ group. Finally, the decay widths
for decays into SM fermions read:
\begin{equation} \label{scalarfermiondecay}
  \Gamma_{\Phi \rightarrow s f \bar f} = \frac{C^2_f\,m^3_{\Phi}}
  {1536\pi^3 \Lambda^2} \,;
\end{equation}
\begin{equation} \label{fermiondecay}
  \Gamma_{\Phi \rightarrow f\overline{f}} = \frac{C^2_f\,v^2_s\,m_{\Phi}}
  {16\pi \Lambda^2} = \left(\frac{96\pi^2 v^2_s}{m^2_{\Phi}}\right)
  \Gamma_{\Phi \rightarrow s f \bar f}\,.
\end{equation}
Note that according to the second equality in eq.(\ref{fermiondecay})
the three--body decay $\Phi \rightarrow s f \bar f$ dominates over the
two--body decay $\Phi \rightarrow f \bar f$ if
$m_\Phi > 10 \pi v_s \simeq 7.5$ TeV.

In eqs.(\ref{scalardecay}) and (\ref{scalarfermiondecay}), we have
assumed a single scalar degree of freedom $s$, as appropriate if $S$
is the Higgs doublet of the SM and unitary gauge is used. In our
numerical computations, we actually find it more useful to {\em not}
use unitary gauge. Instead, we will treat these decay modes in the
limit of unbroken $SU(2) \times U(1)_Y$, but we will include
$\Phi \rightarrow f \bar f$ two--body decays which are possible only
if $v_s \neq 0$. In this case, the widths (\ref{scalardecay}) and
(\ref{scalarfermiondecay}) have to be multiplied by four.

As justification, note first of all that the longitudinal gauge bosons
only contribute at order $(m_A/m_\Phi)^2$ to $\Phi \rightarrow A A$
decays, due to the structure of the $\Phi A A$ vertex. In contrast,
the emission of a longitudinal gauge boson in
$\Phi \rightarrow f \bar f$ decays is enhanced by a factor of order
$E_A/v_s$, which cancels the $v_s$ factor in the $\Phi f \bar f$
coupling; here $E_A$ is the energy of the gauge boson in the $\Phi$
rest frame. The resulting partial width is identical to that for
$\Phi \rightarrow f \bar f G$ decays, where $G$ is the would--be
Goldstone mode ``eaten'' by the longitudinal gauge boson. Finally, in
unitary gauge we would have to include the mixing between $\Phi$ and $s$,
which is a necessary consequence of the $\Phi |S|^2$ coupling if
$\langle S \rangle \neq 0$. This mixing, which is of order
$C_S v_s/m^2_\Phi$, would lead to a matrix element for the decay of
the physical $\Phi$ into two longitudinal gauge bosons of order
$(C_S v_s / m_\Phi^2) g_A m_A (E_A/m_A)^2$, where $g_A$ is the gauge
coupling of $A$. Since $E_A = m_\Phi/2$ and $g_A v_S = 2 m_A$ this
gives a result equal to that for $\Phi \rightarrow G G$ decays, up to
(negligible) terms of order $m_A^2 / m_\Phi^2$. All this is in accord
with the electroweak equivalence theorem \cite{Cornwall:1974km,
  Lee:1977eg}.

In the calculations for the branching ratios, we will assume that these
decay modes, which are accessible at leading order in perturbation
theory, dominate the total decay width of the modulus.  Moreover, we
assume that DM particles can be produced only through higher order
diagrams, as indicated by the bound (\ref{bbound}).

In this section, we compute the corresponding branching ratios in the
framework of simplified models. To that end, we introduce two WIMP
candidates: a Dirac fermion $\chi$ and a complex scalar $\phi$, which
are neutral under $U(1)_{\text{EM}}$. Of course, often only one of
these candidates will be present in a given model. We are interested
in scenarios where only one additional vertex is needed in order to
produce WIMPs in $\Phi$ decays; this will be sufficient for our main
application, the MSSM, which will be discussed in the next section. We
therefore introduce the following renormalizable couplings of the WIMP
candidates to the above list of particles with direct couplings to
$\Phi$:\footnote{We give here the general form of the vertices. Of
course, in general, there will be several different relevant gauge
bosons $A$, scalars $S$ and/or fermions $f$, which may also mix
among themselves. If the WIMP transforms non--trivially under some
non--Abelian gauge group, group generators will need to be
introduced in the corresponding vertices.  None of this alters the
Lorentz structure of the vertices.}
\begin{subequations} \label{l_int}
\begin{equation} \label{l_chi}
  \mathcal{L}_\chi^{\rm int} = -g \overline{\chi} \gamma^\mu(c_V + c_A\gamma^5)
  \chi A_\mu + \alpha s\overline{\chi}(a + b\gamma^5)\chi + \text{h.c.} \,;
\end{equation}
\begin{equation} \label{l_phi}
  \mathcal{L}_\phi^{\rm int} = - g A^\mu(\phi^\ast\partial_\mu\phi
  - \phi\,\partial_\mu\phi^\ast) + \lambda  s|\phi|^2  \,.
\end{equation}
\begin{equation} \label{l_phi_chi}
  \mathcal{L}^{\rm int}_{\phi\chi} = \phi \overline{\chi} (c + d\gamma^5)f
  +\text{h.c.} \,.
\end{equation}
\end{subequations}
All symmetries of our simplified model allow $\Phi \bar \chi \chi$ and
$\Phi \phi^* \phi$ couplings. Since these are renormalizable interactions,
in their presence $\Phi \rightarrow {\rm DM}$ decays could very easily
dominate over the SM modes. Of course, in that case our effective $B_{\Phi
\rightarrow {\rm DM}}$ would be close to $1$, not $\lsim 10^{-4}$, hence
we have to forbid these terms. At the level of simplified models, this is
ad hoc, but we will see in Section \ref{moddecayssusy} how it can be achieved in the context
of supersymmetric models. 

The constants $g$ and $\alpha$ describe the overall strengths of the
couplings of $\chi$ to gauge and scalar bosons; the parametrization of
these couplings is therefore somewhat redundant. Since we take
$m_f = 0$ the $\phi \chi f$ couplings will always appear in the
combination $|c|^2 + |d|^2$, therefore we do not introduce a separate
quantity describing the overall strength of this coupling. The masses
of these dark sector particles are assumed to be greater than those of
$A_\mu$ and $s$. If on--shell decays of these particles into WIMPs
were allowed the bound (\ref{bbound}) on the $\Phi \rightarrow$ WIMP
branching ratio would almost certainly be violated. In fact, we will
continue to neglect the masses of all $A_\mu, \, s$ and $f$, while
keeping $m_\chi$ and $m_\phi$ nonzero. Since this will force the
propagators in diagrams leading to WIMP production in $\Phi$ decay to
be slightly more off--shell than for finite masses, this will slightly
underestimate the branching ratios for these higher order decays, but
the effect will be small over most of parameter space.

In the calculations presented below, the decaying modulus is always
assigned four-momentum $P$, with $P^2 \equiv P \cdot P = m^2_\Phi$,
while the four--momenta of the final state particles are denoted $p_i$
with $i$ running up to 3 or 4. The four--momentum running through the
single propagator in a given diagram is denoted by $k$, which can
always be expressed as the sum of two final--state four--momenta. We
will typically work in the rest frame of the decaying modulus.

In the following three subsections, we discuss higher order decays of
the modulus into our WIMP candidates via the couplings $C_S,\; C_g$
and $C_f$, respectively. In all cases, we computed the squared,
spin--summed matrix elements both ``manually'' and with the help of
FeynCalc \cite{FeynCalc1, FeynCalc2}. We performed those phase space
integrals analytically that yield simple functions, and also provide
analytical expressions for the totally integrated widths that are
accurate in the limit $m^2_\chi,\; m^2_\phi \ll m^2_\Phi$. In the
final subsection, we discuss upper bounds on the mediator couplings
that can be derived from the requirement (\ref{bbound}).

\subsection{Decays into Light Scalars}

\begin{figure} [h]
\centering
\includegraphics[scale=0.9]{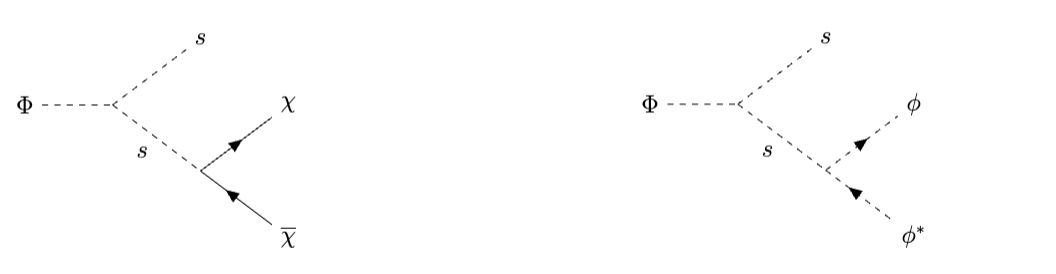}
  \caption{Three--body decays of the modulus from the coupling
    to two scalars. This and all other diagrams in this work have been
    generated using TikZ-Feynman \cite{TikZFeyn}.}
  \label{fig:HOPhi2h}
\end{figure}

We begin with higher order decays arising from the vertex involving
two SM--like scalars, as shown in Figs.~\ref{fig:HOPhi2h}. We denote
the 4--momentum of $s$ by $p_1$, while $p_2$ and $p_3$ are the
4--momenta of the DM particle and antiparticle, respectively. The
squared, summed matrix elements for the production of two fermionic or
two scalar WIMPs together with an $s$ particle can then be written as:
\begin{subequations} \label{eq:s}
  \begin{equation} \label{eq:schichi}
    |\mathcal{M}(\Phi \rightarrow s \chi \bar \chi)|^2 = \frac{C^2_{S}\alpha^2}
  {k^4} \,4[(|a|^2 + |b|^2)\, (p_2 \cdot p_3) - (|a|^2 -|b|^2) m^2_{\chi}]
    \; ;
  \end{equation}
\begin{equation} \label{eq:sphiphi}
  |\mathcal{M}(\Phi \rightarrow s \phi \phi^{\ast})|^2 = \frac{C^2_S \lambda^2}
  {k^4}\;.
\end{equation}
\end{subequations}
Referring to Appendix \ref{phasespacefac}, we factorize the
three--body phase space into a single--particle phase space for $s$
and a two--body phase space for the subsystem of the dark sector
particles, with an additional integral over the squared invariant mass
of that subsystem, $m_X^2 = (p_2 + p_3)^2$. Only this last integral is
somewhat nontrivial, leading to the following expressions for the
corresponding total widths:
\begin{subequations} \label{HOScalardecay}
\begin{equation} \label{eq:Gschichi}
  \Gamma_{\Phi \rightarrow s \chi  \overline{\chi}} = \left( \frac {C^2_S}
    {32\pi  m_\Phi} \right) \frac{\alpha^2}{4\pi^2} \int_{4m^2_\chi}^{m^2_\Phi}
  \left[ \frac{(|a|^2 + |b|^2)m^2_X  - 4 |a|^2 m^2_\chi} {m^4_X} \right]
  \left( 1 - \frac {m^2_X} {m^2_\Phi} \right) \sqrt{1 - \frac{4m^2_\chi}
    {m^2_{X}}} \, dm^2_X \;;
\end{equation}
\begin{equation} \label{eq:Gsphiphi}
  \Gamma_{\Phi \rightarrow s \phi \phi^{\ast}} = \left(\frac{C^2_S}{32\pi  m_\Phi}
  \right) \frac {\lambda^2} {8\pi^2 m^2_\Phi} \int_{4 m^2_\phi}^{m^2_\Phi}
  \frac {m^2_\Phi} {m^4_X} \left( 1 - \frac {m^2_X} {m^2_\Phi} \right)
  \sqrt{1 - \frac {4m^2_\phi} {m^2_X} } \,dm^2_X \;.
\end{equation}
\end{subequations}
The factor in parentheses appearing in both eq.(\ref{eq:Gschichi}) and
eq.(\ref{eq:Gsphiphi}) is the total width for the leading--order decay
$\Phi \rightarrow s s$, see eq.(\ref{scalardecay}). If $C_S \Lambda / m_\Phi^2
\gg C_f,\; C_g$, so that $\Phi \rightarrow s s$ decays determine the total
width of $\Phi$ to leading order, the branching ratios for $\Phi \rightarrow
s \chi \bar \chi$ and $\Phi \rightarrow s \phi \phi^\star$ decays are therefore
simply given by eqs.(\ref{HOScalardecay}) with this factor omitted.
The branching ratios can then be made small by reducing the coupling factor
immediately in front of the remaining integral, and/or by choosing WIMP
masses near $m_\Phi/2$ so that the phase space integrals become small.

More often one is interested in the opposite regime, where the dark
sector masses are much smaller than the modulus mass,
i.e. $m^2_{\chi,\phi} \ll m^2_\Phi$. In order to incorporate this
consistently, we expand the square root appearing in
eqs.\eqref{HOScalardecay} to first order:
\begin{equation*}
\sqrt{1 - \frac {4m^2_{\chi,\phi}} {m^2_X} } = 1 - \frac {2m^2_{\chi,\phi}}{m^2_X}
  + \mathcal{O} \left(\left(\frac{4m^2_{\chi,\phi}}{m^2_X}\right)^{2}\right) \,.
\end{equation*}
We consistently keep the masses of the dark sector particles only
where they are needed to regulate infrared (IR) divergences that would
otherwise result from the integral over the virtuality of the $s$
propagator, but neglect them everywhere else. The remaining integrals
can then easily be expressed as elementary functions:
\begin{subequations} \label{HOScalarapprox}
\begin{equation} \label{ap1}
  \Gamma_{\Phi \rightarrow s \chi  \overline{\chi}} \approx \left( \frac {C^2_S}
    {32\pi  m_\Phi}\right) \frac {\alpha^2 } {4\pi^2} \left[
    (|a|^2+|b|^2) \log \left( \frac {m^2_\Phi} {4m^2_\chi} \right)
    - \frac {(9|a|^2+6|b|^2)} {4} \right] \;;
\end{equation}
\begin{equation} \label{ap2}
  \Gamma_{\Phi \rightarrow s \phi \phi^{\ast}} \approx \left( \frac {C^2_S}
    {32\pi m_\Phi} \right) \frac {\lambda^2} {8\pi^2 m^2_\Phi} \left[
    \frac {m^2_\Phi} {6m^2_{\phi}} - \log \left( \frac {m^2_\Phi} {m^2_\phi}
    \right) + 1 \right] \;.
\end{equation}
\end{subequations}
Note that the decay of the modulus to two dark scalars is formally
more badly behaved, due to the presence of a $1/m^2_\phi$ term. However,
physically one does not expect $\lambda^2 \gg m_\phi^2$, since this
dimensionful coupling contributes to the $\phi$ mass at the one--loop
level; moreover, too large values of $|\lambda|$ lead to the
occurrence of undesirable minima of the scalar potential where
$\langle \phi \rangle \neq 0$. If $\lambda^2 \sim m_\phi^2$ the
branching ratio into $s \phi \phi^*$ does not show any enhancement for
small WIMP mass, in contrast to that for the $s \chi \bar\chi$ final
state.

\begin{figure}[t] 
\centering
 \includegraphics[width=0.75\textwidth]{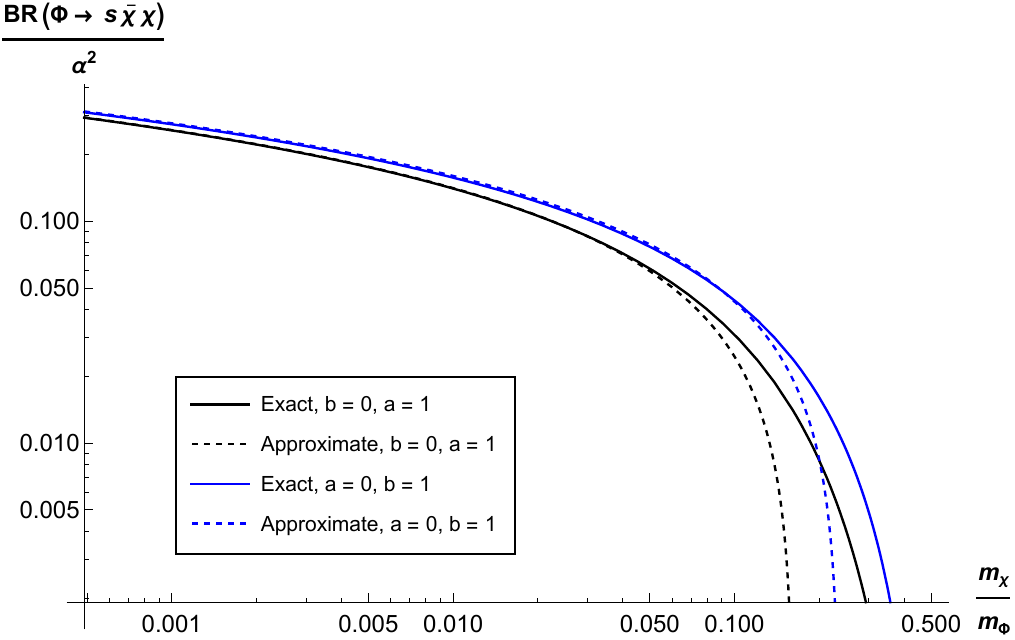}
 \caption{Comparisons between the numerically computed decay
   width (\ref{eq:Gschichi}) and approximate analytic expression
   (\ref{ap1}) for the modulus decay to two fermionic WIMPs and a light
   scalar. We consider the cases of a pure scalar coupling ($b=0$)
   and a pure pseudo--scalar coupling ($a=0$). In both cases, our
   approximations track the exact results very well as long as
   $m_\chi \lsim 0.1 m_\Phi$. }
 \label{fig:schichi}
\end{figure}
\begin{figure}[h]
\centering
\includegraphics[width=0.75\textwidth]{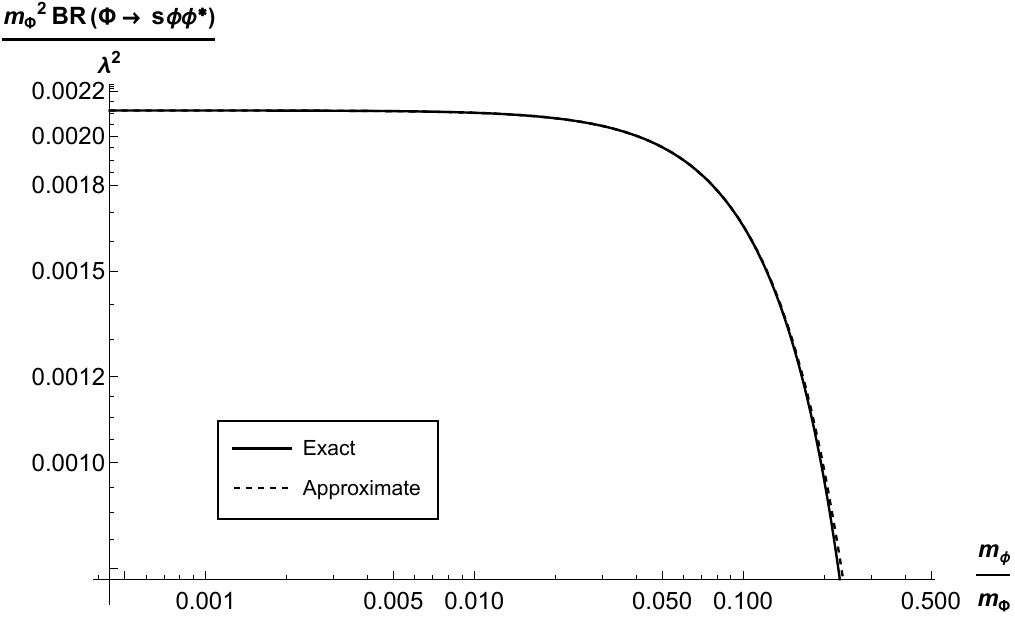}
\caption{Comparison between the numerically computed decay width
  (\ref{eq:Gsphiphi}) and approximate analytic expression (\ref{ap2})
  for the modulus decay to two scalar WIMPs and a light scalar. The
  approximate expression tracks the numerical result extremely well as
  long as $m_\phi \lsim 0.3 m_\Phi$; note in particular the behavior
  $\propto m_\phi^{-2}$ at small WIMP mass.}
\label{fig:sphiphi}
\end{figure}

Figs.~\ref{fig:schichi} and \ref{fig:sphiphi} show numerical results
for the branching ratios for $\Phi \rightarrow s \chi \bar\chi$ and
$\Phi \rightarrow s \phi \phi^\ast$ decays, respectively. For the
fermionic DM particle $\chi$ we show results for purely scalar ($b=0$)
and purely pseudoscalar coupling ($a=0$) to $s$, divided by $\alpha^2$
or, equivalently, for $\alpha = 1$. They agree for
$m_\chi \rightarrow 0$, but the scalar coupling leads to stronger
phase space suppression at larger $m_\chi$; this can easily be
understood from the term in square parentheses in
eq.(\ref{eq:Gschichi}), and corresponds to the stronger phase space
suppression of the decay of a scalar (rather than pseudoscalar)
spin--0 boson into fermion pairs. Similarly, the approximate
expression (\ref{ap1}), shown as dashed lines, begins to deviate from
the exact result computed from eq.(\ref{eq:Gschichi}) somewhat earlier
for purely scalar coupling.

The branching ratio for $\Phi \rightarrow s \phi \phi^\ast$ is shown in
Fig.~\ref{fig:sphiphi} for $\lambda^2 = m_\phi^2$. As already remarked,
for this natural choice of $\lambda$ the branching ratio is not enhanced
as $m_\phi \rightarrow 0$, in contrast to the case of fermionic DM particle
discussed in the previous paragraph. The approximation (\ref{ap2}) works
very well for $m_\phi \lsim 0.3 m_\Phi$ but becomes completely unphysical
close to threshold.

\subsection{Decays into Gauge Bosons}

\begin{figure} [b]
\centering
\includegraphics[scale=0.9]{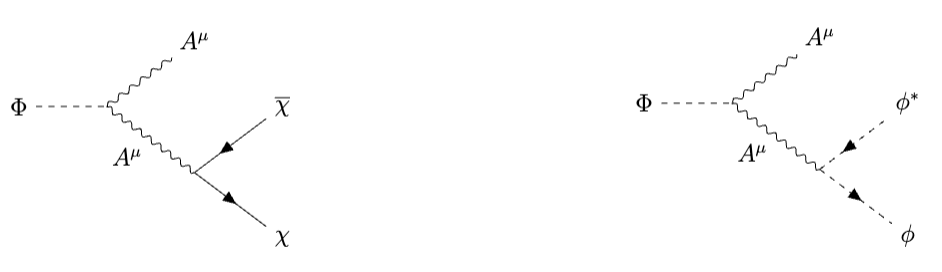}
  \caption{Three--body decays of the modulus from the gauge sector.}
  \label{fig:HOGaugedecay}
\end{figure}

We now turn to three--body decays caused by the coupling $C_g$.  The
relevant diagrams are shown in Fig.~\ref{fig:HOGaugedecay}. We again
begin by calculating the squared and spin--summed matrix elements for
these processes. Denoting the 4--momenta of the two DM particles and
of the on--shell gauge boson with $p_1, \, p_2$ and $p_3$, respectively,
we obtain:
\begin{subequations}
\begin{align}
  |\mathcal{M}(\Phi \rightarrow A\chi \bar\chi)|^2 = \frac{8\,C^2_g g^2}
  {\Lambda^2} \frac {1} {k^4} \{ \nonumber
  & (|c_V|^2 + |c_A|^2) [ (k \cdot p_3) ( (k \cdot p_2) (p_1 \cdot p_3)
   + (k \cdot p_1)(p_2 \cdot p_3) ) \\&- k^2 (p_3 \cdot p_2)(p_3 \cdot p_1) ]
  +(|c_V|^2 - |c_A|^2) m^2_\chi (k\cdot p_3)^2 \}
  \,;
\end{align}
\begin{equation}
  |\mathcal{M}(\Phi \rightarrow A\phi \phi^\ast)|^2 = \frac {C^2_g g^2}
  {\Lambda^2} \frac{1}{k^4} [ -(k\cdot p_3)^2 (p_1 - p_2)^2
  + 2(k\cdot p_3) (k \cdot p_1 - k \cdot p_2) (p_1 \cdot p_3 - p_2 \cdot p_3)
  - k^2 (p_1 \cdot p_3 - p_2 \cdot p_3)^2]\,.
\end{equation}
\end{subequations}
Here we have assumed a single gauge boson, i.e. a $U(1)$ gauge group. In general,
a group factor will have to be included. For example, if $\chi$ and $\phi$
transform as fundamental representations of a non--Abelian $SU(N)$ group,
the result needs to be multiplied with $(N^2-1)/2$.

In order to make progress, we now go to the rest frame of the modulus
and adopt the scaled variable approach \cite{Barger:1987book} to
describe the three--body phase space. Here one defines
$x_i \equiv 2 E_i / m_\Phi$ so that $\sum_{i=1}^3 x_i = 2$. After some
simplification, this results in the following expressions for the
squared and summed matrix elements:
\begin{subequations}
\begin{equation}
|\mathcal{M}(\Phi \rightarrow A\chi \bar\chi)|^2 = \frac{C^2_g g^2 m^2_\Phi}
{\Lambda^2} \frac {2 (|c_V|^2 - |c_A|^2) \mu_\chi x^2_3
  + (|c_V|^2 + |c_A|^2) (1-x_3) \left[x^2_3 - 2x_3(1-x_2) + 2(1-x_2)^2 \right] }
{(1-x_3)^2} \,;
\end{equation}
\begin{equation}
  |\mathcal{M}(\Phi \rightarrow A\phi \phi^\ast)|^2 = \frac{C^2_g g^2 m^2_\Phi}
  {4 \Lambda^2} \frac{ x^2_3 (1 - 4 \mu_\phi - x_3 )
    - (1 - x_3) (2 - 2x_2 - x_3)^2} { (1-x_3)^2 }\,,
\end{equation}
\end{subequations}
where $\mu_{\chi,\phi} = m^2_{\chi,\phi}/m^2_\Phi$. The decay widths
are hence given by
\begin{subequations}
\begin{equation}
\begin{split}
  \Gamma_{\Phi \rightarrow A\chi \bar\chi} &= \left( \frac {C^2_g m^3_\Phi }
    { 64 \pi \Lambda^2 }\right) \frac {g^2} {4\pi^2} \\& \cdot \int \int
  \frac {2(|c_V|^2 - |c_A|^2) \mu_\chi x^2_3 + (|c_V|^2 + |c_A|^2) (1-x_3)
    \left[ x^2_3 - 2x_3(1-x_2) + 2(1-x_2)^2 \right] } { (1-x_3)^2}
  dx_2 \, dx_3 \,;
\end{split}
\end{equation}
\begin{equation}
  \Gamma_{\Phi \rightarrow A\phi \phi^\ast} = \left( \frac {C^2_g m^3_\Phi }
    {64 \pi \Lambda^2} \right) \frac {g^2} {16\pi^2} \int \int
  \frac {x^2_3 (1 - 4\mu_{\phi} - x_3) - (1-x_3) (2 - 2x_2 - x_3)^2}
  {(1-x_3)^2}  dx_2 \,dx_3 \,.
\end{equation}
\end{subequations}

\begin{figure}[t]
\centering
\includegraphics[width=0.75\textwidth]{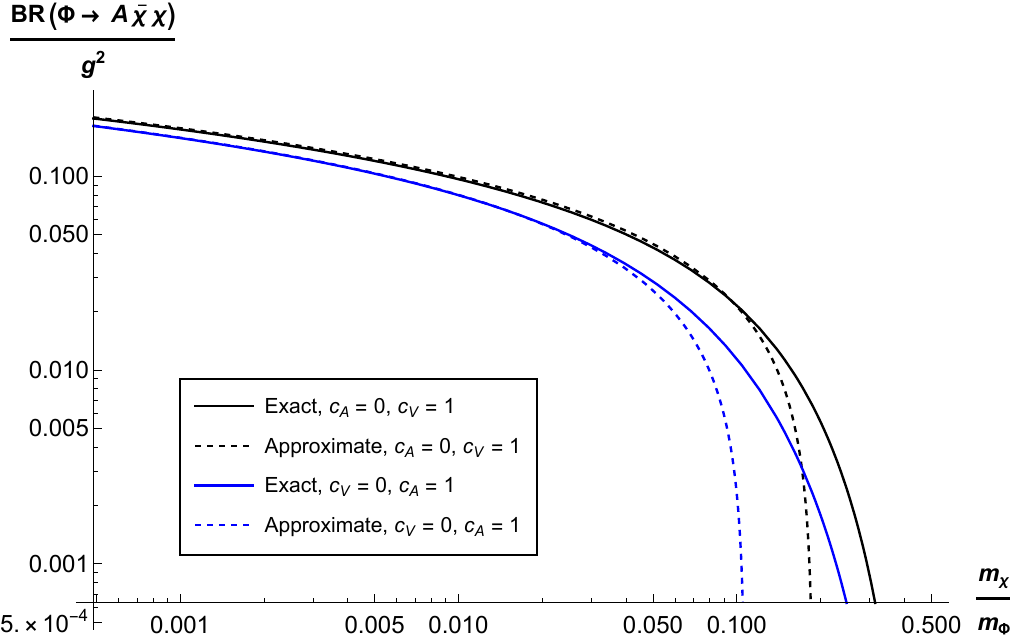}
\caption{The branching ratio of three--body decays into a gauge boson
  plus fermionic DM. Solid curves show exact results according to
  eq.(\ref{HOgaugedecay}a), for pure vector coupling ($c_A = 0$, black)
  and a pure axial vector coupling ($c_V = 0$, blue). Dashed curves
  of the same color show the corresponding results of the
  approximation (\ref{ap3}). The agreement is good in both cases, as
  long as $m_\chi \lsim 0.05 m_\Phi$.}
\label{fig:Achichi}
\end{figure}

The limits of integration are
\begin{subequations} \label{limits}
\begin{equation}
  \frac{1}{2} \left[ (2-x_3) - x_3 \sqrt{1 - \frac{4\mu_{\chi,\phi}} {1-x_3} }
  \right] \leq  x_2 \leq \frac{1}{2} \left[ (2-x_3) + x_3
    \sqrt{ 1 - \frac{4\mu_{\chi,\phi}} {1-x_3} } \right] \,;
\end{equation}
\begin{equation}
 0 \leq\, x_3 \leq 1-4\mu_{\chi,\phi}\,.
\end{equation}
\end{subequations}
The integral over $x_2$ is readily evaluated:
\begin{subequations} \label{HOgaugedecay}
\begin{equation}
  \Gamma_{\Phi \rightarrow A\chi \bar\chi} = \left( \frac {C^2_g m^3_\Phi }
    {64 \pi \Lambda^2} \right) \frac {g^2} {4\pi^2} \int^{1-4\mu_{\chi}}_0
  \frac {2x^3_3} {3} \, \sqrt{1 - \frac {4 \mu_{\chi}} {1-x_3} }
  \frac {(|c_V|^2 + |c_A|^2)(1-x_3) + 2 (|c_V|^2 - 2|c_A|^2) \mu_{\chi}}
  {(1-x_3)^2}\, dx_3 \,;
\end{equation}
\begin{equation}
  \Gamma_{\Phi \rightarrow A\phi \phi^\ast} = \left( \frac {C^2_g m^3_{\Phi}}
    {64 \pi \Lambda^2} \right) \frac {g^2} {16\pi^2} \int^{1-4\mu_{\phi}}_0
  \frac {2x^3_3} {3(1-x_3)} \, \left( 1 - \frac{4 \mu_{\phi}}{1-x_3}
  \right)^{3/2} \,dx_3 \,.
\end{equation}
\end{subequations}
We have again factorized the partial decay widths into a factor of
the width for $\Phi \rightarrow A A$ two--body decays, an additional
coupling factor involving the gauge coupling $g$, and a phase space
integral that becomes much smaller than 1 only if the mass of the DM
particles approaches $m_\Phi/2$.

\begin{figure}[H]
\centering
\includegraphics[width=0.75\textwidth]{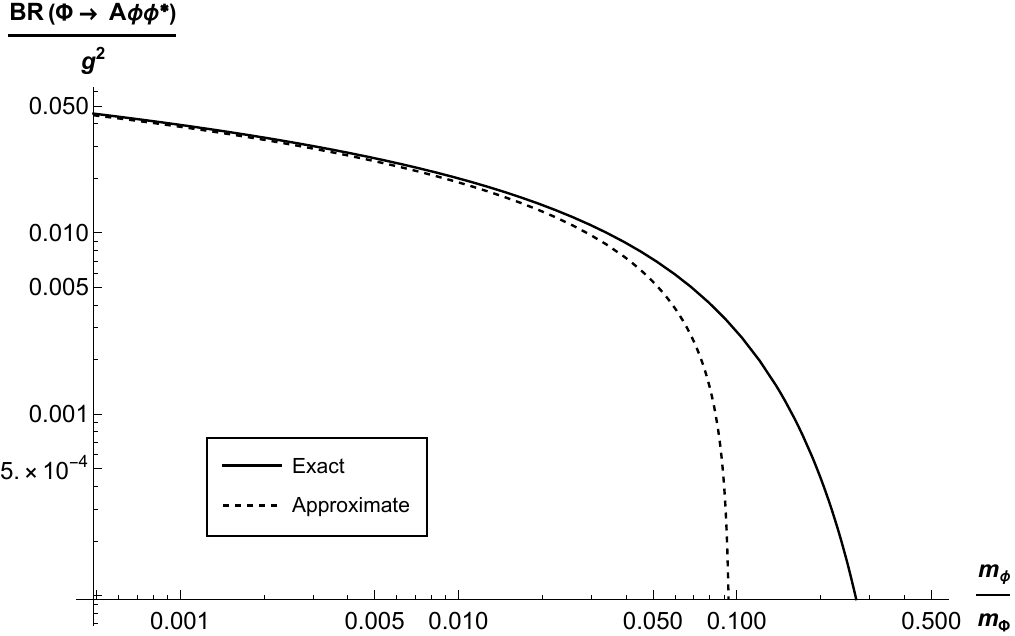}
\caption{The branching ratio of three--body decays into a gauge boson
  plus scalar DM. The solid curve shows exact results according to
  eq.(\ref{HOgaugedecay}b), while the dashed curve shows the
  corresponding results of the approximation (\ref{ap4}). The
  agreement is good as long as $m_\phi \lsim 0.05 m_\Phi$.}
\label{fig:Aphiphi}
\end{figure}

In the phenomenologically most interesting limit 
$m^2_{\chi,\phi} \ll m^2_{\Phi}$, the integrals over $x_3$ can also be
evaluated analytically:
\begin{subequations} \label{apgauge}
\begin{equation} \label{ap3}
 \Gamma_{\Phi \rightarrow A\chi \bar\chi} \approx \left( \frac {C^2_g m^3_\Phi}
   {64\pi \Lambda^2} \right) \frac {g^2} {4\pi^2} \left[
   \frac {2}{3} (|c_V|^2 + |c_A|^2) \log \left( \frac {m^2_\Phi}
     {4m^2_\chi}\right) - \frac{ (47|c_V|^2 + 74|c_A|^2) } {36} \right]  \,;
\end{equation}
\begin{equation} \label{ap4}
  \Gamma_{\Phi \rightarrow A\phi \phi^\ast} \approx \left( \frac {C^2_g m^3_\Phi}
    {64\pi \Lambda^2} \right) \frac {g^2} {16\pi^2} \left[ \frac {2} {3}
    \log \left( \frac {m^2_\Phi} {4m^2_\phi} \right) - \frac{20}{9}\right]\,.
\end{equation}
\end{subequations}
Results for the branching ratios for three--body decays into a gauge
boson plus fermionic or scalar DM are shown in Figs.~\ref{fig:Achichi}
and \ref{fig:Aphiphi}, respectively. Here it has been assumed that the
total decay width is dominated by decays into two gauge bosons, and
the branching ratio has been divided by $g^2$ in order to make it
independent of the gauge coupling strength. In case of fermionic DM we
observe stronger phase space suppression for axial vector coupling,
which shows behavior similar to that of scalar DM. In these two cases,
the analytical approximation works well for DM mass up to about
$0.05 \, m_\Phi$, whereas for pure vector coupling the approximation
(\ref{ap3}) works well for $m_\chi \lsim 0.1 \, m_\Phi$. 

\subsection{Decays into Light Fermions}

We next examine the higher order decays arising from the coupling
$C_f$. Recall that already at leading order it gives rise to two--body
decays $\Phi \rightarrow f \bar f$ as well as three--body decays
$\Phi \rightarrow f \bar f s$, the latter being dominant for
$m_\Phi \gsim 7.5$ TeV. Going one order higher in perturbation theory
therefore allows three-- and four--body decays, which we discuss in
turn.

\begin{figure}[H]
\centering
\includegraphics[scale=0.9]{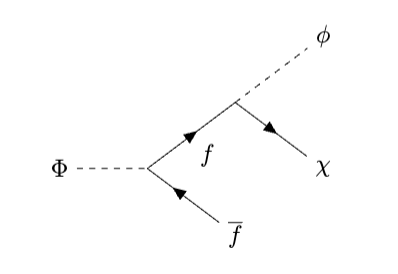}
   \caption{The three--body decay arising from the coupling of the
     modulus to an SM--like fermion, ultimately again leading to a
     final state with two DM particles.}
\label{fig:C_f_3}
\end{figure}

We begin with three--body decays involving the $f \phi \chi$ couplings
introduced in eq.(\ref{l_phi_chi}),
$\Phi \rightarrow \phi \chi \bar f$. The corresponding Feynman diagram
is shown in Fig.~\ref{fig:C_f_3}. Note that we now have both potential
DM candidates in the final state. Since $f$ is (almost) massless, the
heavier of $\phi$ and $\chi$ will be able to decay into the lighter
one plus $f$ or $\bar f$ via the same coupling, so effectively this
decay also produces two DM particles.

Denoting the four--momenta of the light fermion, $\phi$ and $\chi$ by
$p_1, \, p_2$ and $p_3$, respectively, the summed and squared matrix
element for this process is:
\begin{equation} \label{ffphi}
  |\mathcal{M}(\Phi \rightarrow \bar{f}\chi \phi)|^2 =
  \frac{2 C^2_f v^2_s } {\Lambda^2} (|c|^2 + |d|^2)
  \frac {2 (k\cdot p_2) (k\cdot p_3) - k^2 (p_2 \cdot p_3) } {k^4} \,;
\end{equation}
recall that $k = p_2 + p_3$ is the four--momentum in the off--shell
propagator. We use the technique of scaled variables $x_i$ to rewrite
this as
\begin{equation*}
  |\mathcal{M}(\Phi \rightarrow \bar{f}\chi \phi)|^2 =
  \frac{C^2_f v^2_s } {\Lambda^2} (|c|^2 + |d|^2)
  \frac { (1 - x_3 + \mu_\chi - \mu_\phi) x_3 - (1 - x_3)
    ( 1 - x_1 + \mu_\phi - \mu_\chi ) } { (1-x_3)^2 } \,,
\end{equation*}
where again $\mu_{\chi,\phi} = m^2_{\chi,\phi}/m^2_\Phi$. We thus
obtain the following expression for the decay width:
\begin{equation} \label{eq:C_f_3}
  \Gamma_{\Phi \rightarrow \bar{f}\chi \phi} = \left( \frac {C^2_f v^2_s m_\Phi }
    { 16 \pi \Lambda^2} \right) \frac {(|c|^2 + |d|^2)} {16 \pi^2}
  \int dx_3 \frac{ x^2_3 ( 1 - x_3 + \mu_\chi - \mu_\phi ) } { 2 (1-x_3)^3 }
  \uplambda^{1/2}( 1 - x_3,\, \mu_\phi,\,\mu_\chi)\,,
\end{equation}
where the kinematic ``triangle'' function is as usually defined:
$\uplambda(a,b,c) = (a - b - c)^2 - 4 bc$. The limits of the remaining
$x_3$ integral are:
\begin{equation*}
0 \leq x_3 \leq 1 - \mu_\phi - \mu_\chi - 2 \sqrt{ \mu_\phi \mu_\chi }\,.
\end{equation*}
By expanding the square--root of the triangle function up to second order
in the $\mu_i$ we find the following approximate analytical result, valid
for $(m_\phi + m_\chi)^2 \ll m^2_\Phi$:
\begin{equation} \label{ap5}
  \Gamma_{\Phi \rightarrow \bar f \chi \phi} \approx \left(
       \frac {C^2_f v^2_s m_\Phi} { 16 \pi \Lambda^2} \right)
       \frac { (|c|^2 + |d|^2) } {32 \pi^2} 
  \bigg\{ 2 \log \left( \frac { m_\Phi } { \Sigma_{\phi\chi} }\right)
    - \frac {3}{2} - \frac {2 m^2_\phi} {\Sigma_{\phi\chi}^2}
  + \frac { m^4_\phi - m^4_\chi - 2 m^2_\chi m^2_\phi} {2 \Sigma_{\phi\chi}^4}
    \bigg\}\,,
\end{equation}
where we have introduced
\begin{equation} \label{eq:Sigma}
  \Sigma_{\phi\chi} = m_\phi + m_\chi\,.
\end{equation}

\begin{figure}[t]
\centering
\includegraphics[width=0.75\textwidth]{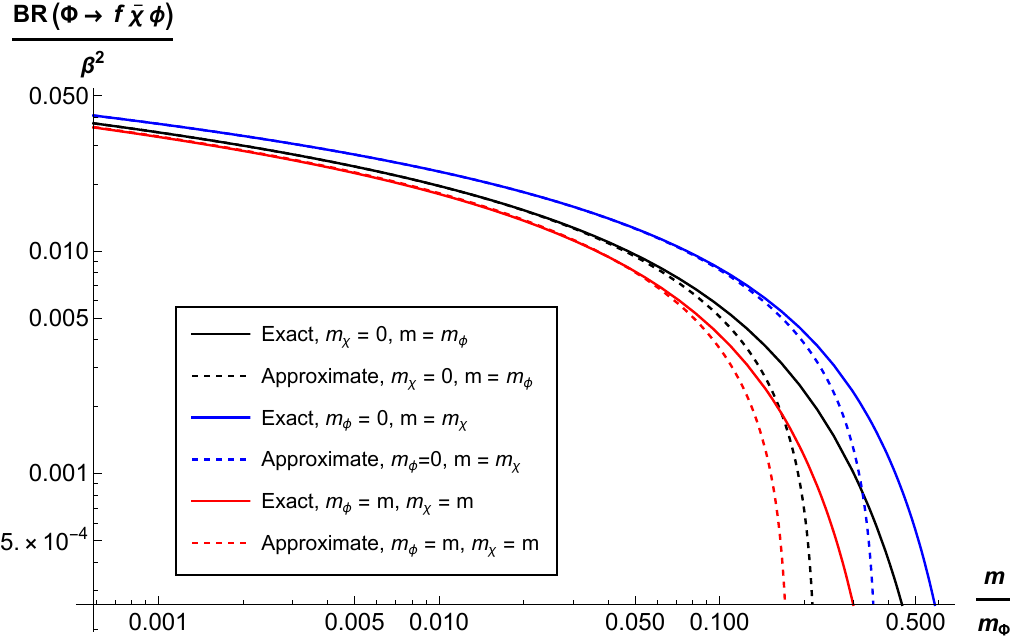}
\caption{Scaled branching ratio for
  $\Phi \rightarrow \phi \chi \bar f$ decays; a factor of 2 has been
  included for the charge conjugate final state. The solid and dashed
  curves of a given color show exact results and the numerical
  approximation (\ref{ap5}), respectively. We consider cases where one
  of the dark sector particles is (effectively) massless (blue,
  black), and when they are mass degenerate (red). In all cases the
  approximation works well as long as the mass of the heavier DM
  candidate satisfies $m \lsim 0.1 m_\Phi$}
\label{fig:fchiphi1}
\end{figure}

Numerical results for the corresponding branching ratios are shown in
Fig.~\ref{fig:fchiphi1}. We see that the approximation (\ref{ap5})
works very well for $m_\phi + m_\chi \lsim 0.1\, m_\Phi$; this is true even
for $m_\phi m_\chi \neq 0$ where eq.(\ref{ap5}) is not quite exact even for
very small DM masses.

\begin{figure} [t]
\centering
\includegraphics[scale=0.9]{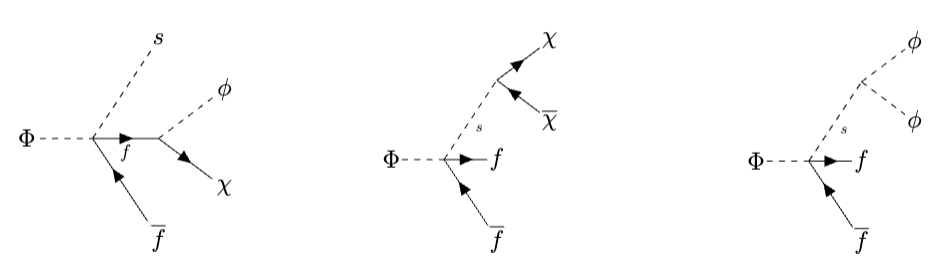}
\caption{Feynman diagrams for four--body decays of $\Phi$ involving the
  $\Phi s \bar f f$ coupling $C_f$.}
\label{4bodydecay}
\end{figure}

We now turn to four--body decays of the modulus involving the coupling
$C_f$. There are three such processes, see Fig.~\ref{4bodydecay}:
$\Phi \rightarrow s(p_1) \bar \phi(p_2) \chi (p_3) \bar f (p_4)$ (left
diagram),
$\Phi \rightarrow \chi (p_1) \bar \chi (p_2) f (p_3) \bar f(p_4)$
(center), and
$\Phi \rightarrow \phi (p_1) \phi^* (p_2) f (p_3) \bar f (p_4)$
(right).\footnote{The $f \bar f \chi \bar \chi$ and
  $f \bar f \phi \phi^*$ final states also receive contributions where
  the $s$ propagator attaches to $f$ or $\bar f$, rather than to the
  $\Phi f \bar f$ vertex. However, these contributions, as well as
  similar contributions with a gauge propagator, are of higher order
  in coupling constants; they contribute at the same order as
  $f \bar f s \chi \bar \chi$ and $f \bar f s \phi \phi^*$ five--body
  final states. We therefore ignore them here.} In the first kind of
decay the charge conjugate decay,
$\Phi \rightarrow s f \bar \chi \phi^*$, has the same partial width
but has a physically distinct final state and should thus be included
in the calculation of the total four--body branching ratio. The
corresponding squared and spin--summed matrix elements are:
\begin{subequations}
\begin{equation} \label{ffphiS}
  |\mathcal{M}(\Phi \rightarrow \bar{f}\chi \phi s)|^2 = \frac {2 C^2_f }
  {\Lambda^2} (|c|^2 + |d|^2) \frac { 2 (k \cdot p_3) (k \cdot p_4)
  - k^2 (p_3 \cdot p_4) } {k^4} \,,
\end{equation}
\begin{equation} \label{ffchi}
  |\mathcal{M}(\Phi \rightarrow \bar f f \chi  \overline{\chi}|^2 =
  \frac {4 C^2_f \alpha^2} {\Lambda^2} \frac {(|a|^2 + |b|^2) (p_1 \cdot p_2)
    - (|a|^2 -|b|^2) m^2_\chi } {k^4} \left( p_3 \cdot p_4 \right) \,,
\end{equation}
\begin{equation} \label{ffphiphi}
  |\mathcal{M}(\Phi \rightarrow \bar f f \phi^\ast \phi|^2 =
  \frac {2 C^2_f \lambda^2} {\Lambda^2} \frac { p_3 \cdot p_4 } {k^4}\,.
\end{equation}
\end{subequations}

\begin{figure}[t]
\centering
\includegraphics[width=0.75\textwidth]{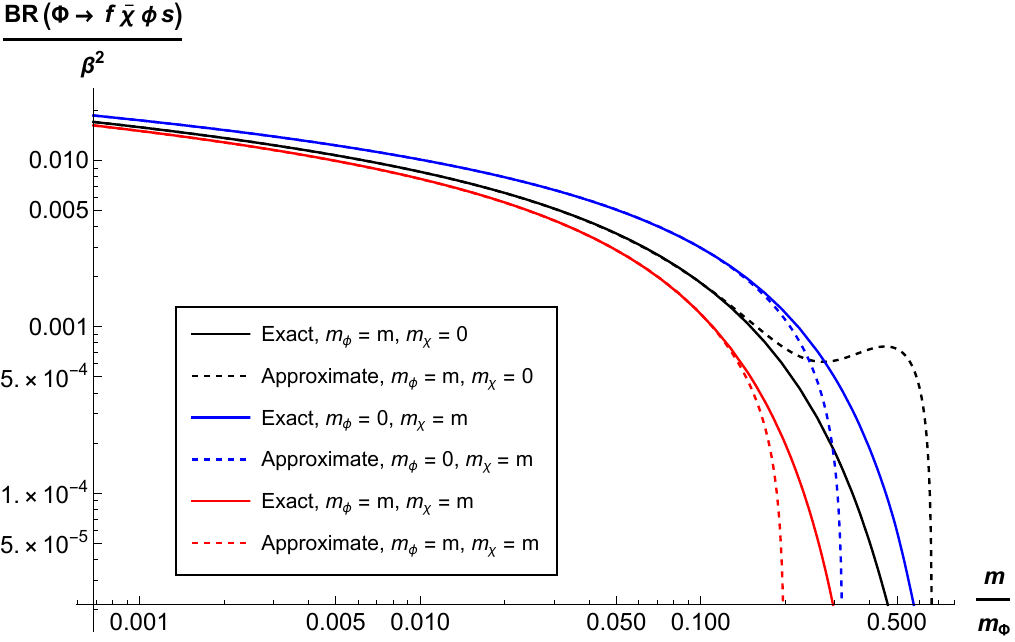}
\caption{Comparisons between the exact branching ratio for
  $\Phi \rightarrow s \bar f \phi \chi$ decays, computed from
  eq.(\ref{eq:sfcp}), and our approximate expression (\ref{ap6}); a
  factor of two has been included to account for the charge conjugate
  final state. We consider here scenarios where one of the dark sector
  particles is massless, or both have the same mass.}
\label{fig:fchiphis1}
\end{figure}

\begin{figure}[H]
\centering
\includegraphics[width=0.75\textwidth]{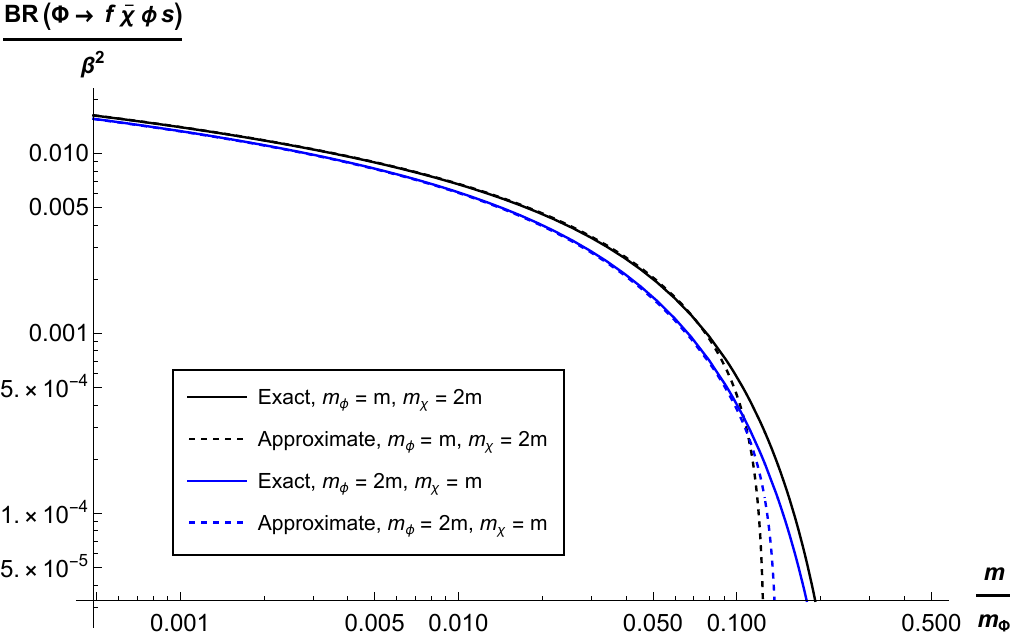}
\caption{As in Fig.~\ref{fig:fchiphis1}, but for scenarios where $m_\chi$
and $m_\phi$ differ by a factor of 2.}
\label{fig:fchiphis2}
\end{figure}

We tackle the four--body phase space by factorization, see Appendix
A. We begin with $\Phi \rightarrow s \bar f \chi \phi$ decays.
Eq.(\ref{ffphiS}) shows that the most convenient subsystem includes
$\phi$, $\chi$ and $\bar f$, so we employ the factorization
(\ref{4body31}). To this end, we define $m^2_X = (p_2 + p_3 + p_4)^2$
and treat this subsystem using the technique of scaled variables,
defining $x_i = 2E_i / m_X$ and $\mu_{\chi, \phi} = m^2_{\chi, \phi}/m^2_X$.
This allows us to rewrite the squared and summed matrix element for
this process as
\begin{equation}
  |\mathcal{M}(\Phi \rightarrow \bar{f}\chi \phi s)|^2{} = \frac {C^2_f }
  {\Lambda^2} (|c|^2 + |d|^2) \frac{ (1 + \mu_\chi - \mu_\phi - x_4) x_4
    - ( 1 - x_4 ) (1 - \mu_\chi + \mu_\phi - x_2) } { (1-x_4)^2} \,.
\end{equation}
The resulting decay width is
\begin{equation}
\begin{split}
  \Gamma_{\Phi \rightarrow \bar f\chi \phi s} &= \left( \frac {C^2_f}
    {1536 \pi^3 \Lambda^2} \right) \frac { (|c|^2 + |d|^2) } {8 \pi^2}
  \frac {3} {m_\Phi} \\
  &\cdot \int dm^2_X m^2_X \left( 1 - \frac {m^2_X} {m^2_\Phi} \right)
  \int \int \frac{ (1 + \mu_\chi - \mu_\phi - x_4) x_4
    - (1 - x_4) (1 - \mu_\chi + \mu_\phi - x_2) } { (1 - x_4)^2} dx_2 dx_4 \,.
\end{split}
\end{equation}
The integration over $x_2$ is rather straightforward, the limits of
integration being
\begin{equation}
  x_2 \lessgtr \frac{ (2-x_4) ( 1 + \mu_\phi - \mu_\chi - x_4) \pm x_4
    \uplambda^{1/2}( 1 - x_4,\mu_\phi,\mu_\chi) } { 2(1-x_4)}\,.
\end{equation}
The final expression for the decay width is thus:
\begin{align} \label{eq:sfcp}
  \Gamma_{\Phi \rightarrow \bar f\chi \phi s} = \left( \frac {C^2_f}
    {1536 \pi^3 \Lambda^2} \right) \frac {3 (|c|^2 + |d|^2) }
  {8 \pi^2 m_\Phi} &\int dm^2_X m^2_X\left( 1 - \frac {m^2_X} {m^2_\Phi} \right)
                      \nonumber \\
  &\int dx_4 \frac {x^2_4 (1 - x_4 + \mu_\chi - \mu_\phi )} { 2(1 - x_4)^3}
  \uplambda^{1/2}(1-x_4,\mu_\phi,\mu_\chi)\,.
\end{align}
The limits for the remaining integrations are
\begin{subequations}
\begin{equation}
  0 \leq\, x_4 \leq 1 - \mu_\phi - \mu_\chi - 2\sqrt{\mu_\phi \, \mu_\chi}\,;
\end{equation}
\begin{equation}
 (m_\chi + m_\phi)^2 \leq m^2_X \leq m^2_\Phi \,.
\end{equation}
\end{subequations}
An accurate approximation for the case $m_{\phi,\,\chi} \ll m_\Phi$ is:
\begin{align} \label{ap6}
  \Gamma_{\Phi \rightarrow \bar{f}\chi \phi s}
  \approx \left(
\frac {C^2_f} {1536 \pi^3 \Lambda^2} \right) \frac {3 m^3_\Phi (|c|^2 + |d|^2) }
  {8 \pi^2}
  &\left\{ \log \left( \frac {m_\Phi}
    {\Sigma_{\phi\chi}} \right) \left( \frac {1} {6}
    + \frac{1.4 m^2_\phi + 0.4 m^2_\chi} {m^2_\Phi}
    +3 \frac {m^2_\phi m^2_\chi} {m^2_\Phi \Sigma_{\phi\chi}^2} \right)
    \right.\nonumber \\
   &\left. - \frac {7} {36} - \frac{ m^2_\phi}{6 \Sigma_{\phi\chi}^2}
     + \frac {m^4_\phi - m^4_\chi - 2m^2_\phi m^2_\chi} {24 \Sigma_{\phi\chi}^4}
     \right\}\,,
\end{align}
where $\Sigma_{\phi\chi}$ had been defined in eq.(\ref{eq:Sigma}).
The coefficients in front of the $m^2_{\phi, \chi}/m^2_\Phi$ terms
multiplying the logarithm have been adjusted to get good results for
$m_\phi = 0$, for $m_\chi = 0$ and for $m_\chi = m_\phi$, for DM mass
up to $20\%$ of the modulus mass, as shown in
Fig.~\ref{fig:fchiphis1}. The results for $m_\phi = 2m_\chi$ and for
$m_\chi = 2m_\phi$ are then not quite as good, but still work very
well for DM masses $\lsim 10\%$ of the modulus mass, as shown in
Fig.~\ref{fig:fchiphis2}. 

We next turn to $\Phi \rightarrow f \bar f \chi \bar\chi$ and
$\Phi \rightarrow f \bar f \phi \phi^*$ decays. Since here the two DM
particles couple to a common ($s$) propagator, we factorize the final
state into two two--body subsystems: one containing $f$ and $\bar f$,
and the other the two DM particles. The invariant masses of these
subsystems are $m^2_Y =(p_1 + p_2)^2 = k^2$ and
$m^2_X =(p_3 + p_4)^2$. The spin--summed and squared matrix elements
can then be written as
\begin{subequations}
\begin{equation}
  |\mathcal{M}(\Phi \rightarrow f \bar f \chi \overline{\chi})|^2 =
  \frac { 2 C^2_f \alpha^2 m^2_X} {\Lambda^2} \frac { (|a|^2 + |b|^2)
    (m^2_Y - 2 m^2_\chi) - 2(|a|^2 -|b|^2) m ^2_\chi} {m^4_Y} \,;
\end{equation}
\begin{equation}
  |\mathcal{M}(\Phi \rightarrow f \bar f \phi \phi^\ast)|^2 =
  \frac {C^2_f \lambda^2} {\Lambda^2} \frac {m^2_X} {m^4_Y} \,.
\end{equation}
\end{subequations}
using eq.(\ref{4body22}) the corresponding partial decay widths are
\begin{subequations}
\begin{equation} \label{eq:ffcc}
\begin{split}
  \Gamma_{\Phi \rightarrow f\bar{f}\chi \overline{\chi}}
  &= \left( \frac {C^2_f} { 1536 \pi^3 \Lambda^2} \right)
  \frac {3\alpha^2} {4 \pi^2 m^3_\Phi} \\
  &\int \int \uplambda^{1/2}(m^2_\Phi, m^2_X, m^2_Y) \sqrt{1 - \frac {4m^2_\chi}
    {m^2_Y} } \frac { (|a|^2 + |b|^2) (m^2_Y - 2 m^2_\chi)
    - 2 (|a|^2 -|b|^2) m ^2_\chi} {m^4_Y} dm^2_Y m^2_X dm^2_X\,,
\end{split}
\end{equation}
\begin{equation} \label{eq:ffpp}
\begin{split}
  \Gamma_{\Phi \rightarrow f\bar{f} \phi \phi^{\ast}}
  &= \left( \frac {C^2_f} {1536 \pi^3 \Lambda^2} \right)
  \frac {3\lambda^2} {8 \pi^2 m^3_\Phi} \int \int
  \uplambda^{1/2}(m^2_\Phi, m^2_X, m^2_Y) \sqrt{1 - \frac{4 m^2_\phi} {m^2_Y} }
  \frac {dm^2_Y} {m^4_Y} m^2_X dm^2_X\,.
\end{split}
\end{equation}
\end{subequations}
The limits of integration are:
\begin{subequations}
\begin{equation}
 4m^2_{\chi,\phi} < m^2_Y < (m_\Phi - m_X)^2 \,;
\end{equation}
\begin{equation}
 0 < m^2_X < (m_\Phi - 2m_{\chi,\phi} )^2 \,.
\end{equation}
\end{subequations}
These decay widths can be approximated by
\begin{subequations}
\begin{align} \label{ap7}
  \Gamma_{\Phi \rightarrow f \bar f \chi \overline{\chi}}
  \approx \left( \frac {C^2_f m^3_\Phi} {1536 \pi^3 \Lambda^2} \right)
  \frac {3 \alpha^2} {4 \pi^2} 
    &\left\{ \log \left( \frac {m^2_\Phi} {m^2_\chi} \right) \left[
    \frac {a^2 + b^2} {6} + (1.16 a^2 + 1.68 b^2) \frac {m^2_\chi} {m^2_\Phi}
    \right]  \right. \nonumber \\
    &\left. + \frac {m^2_\chi} {m^2_\Phi}
      \log^2 \left( \frac {m^2_\Phi} {4m^2_\chi} \right)
      \left[ 2 a^2 + b^2 \right] - 0.997 a^2 - 0.889 b^2\right\} \,;
\end{align}
\begin{equation} \label{ap8}
  \Gamma_{\Phi \rightarrow f \bar f \phi \phi} \approx \left(
    \frac {C^2_f m^3_\Phi} {1536 \pi^3 \Lambda^2} \right) \frac {3\lambda^2}
  {8 \pi^2} \left\{ \frac {1}{36 m^2_\phi} + \frac {187}{6 m^2_\Phi}
    - \frac {166 m_\phi} {5 m^3_\Phi} - \frac {5}{m^2_\Phi}
    \log\left(\frac{m^2_\Phi} {m^2_\phi}
    \right) \left( 1 - 3 \frac {m_\phi} {m_\Phi} \right) \right\}\,.
\end{equation}
\end{subequations}
Here again, some of the coefficients have been fitted numerically. 

\begin{figure}[H]
\centering
\includegraphics[width=0.75\textwidth]{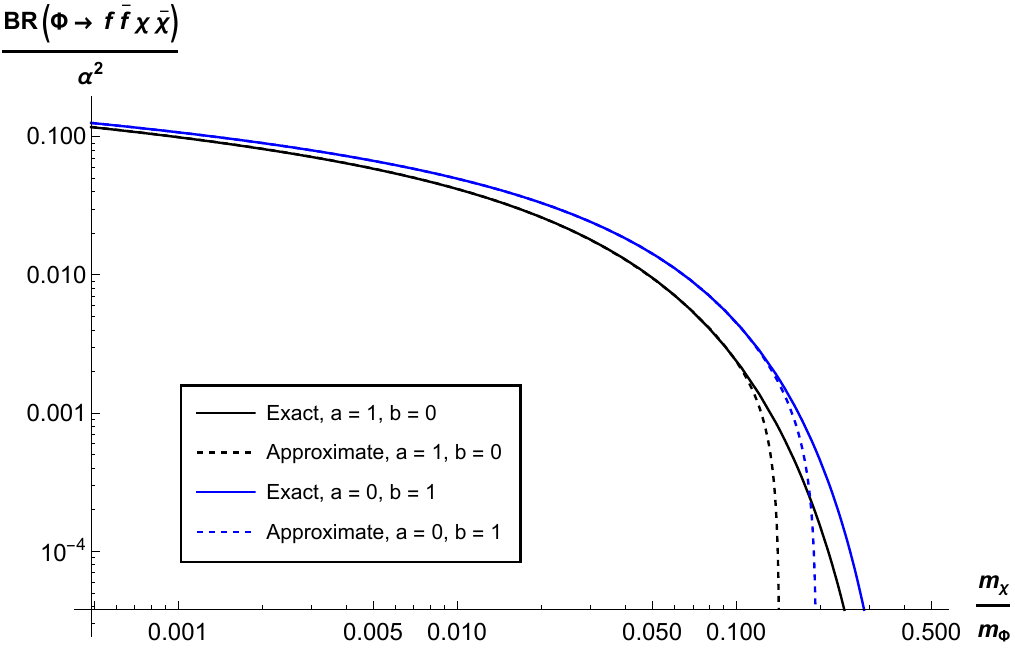}
\caption{The exact branching ratio for
  $\Phi \rightarrow f \bar f \chi \bar \chi$ decays via $s$ exchange,
  computed from eq.(\ref{eq:ffcc}), is compared to the approximation
  (\ref{ap7}). The approximate expression works well for
  $m_\chi \lsim 0.2 m_\Phi \ (0.1 m_\Phi)$ for pure (pseudo--)scalar
  coupling of $\chi$ to $s$.}
\label{fig:ffcc}
\end{figure}

\begin{figure}[H]
\centering
    \includegraphics[width=0.75\textwidth]{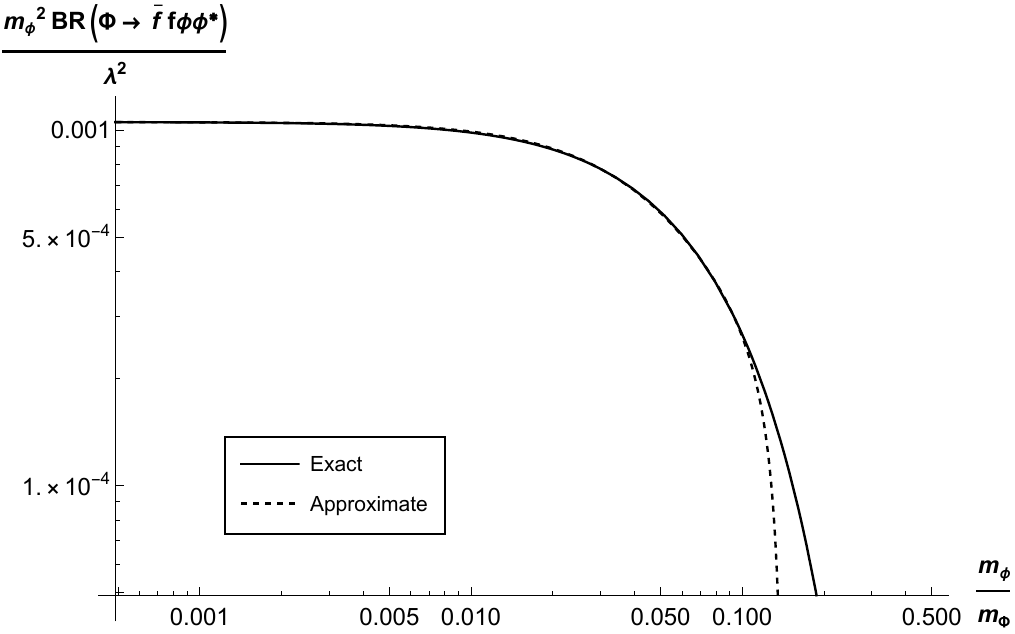}
\caption{The exact branching ratio for
  $\Phi \rightarrow f \bar f \phi \phi^*$ decays, computed from
  eq.(\ref{eq:ffpp}), is compared to the approximation
  (\ref{ap8}). The approximate expression works well for
  $m_\phi \lsim 0.1 m_\Phi$.}
\label{fig:ffpp}
\end{figure}

Figs.~\ref{fig:ffcc} and \ref{fig:ffpp} show some numerical
results. In the case of fermionic DM we again observe stronger kinematical
suppression for scalar $s \chi \bar \chi$ coupling (black lines in
Fig.~\ref{fig:ffcc}) than for pseudoscalar coupling (blue). Moreover,
under the natural assumption that the dimensionful $s \phi \phi^*$
coupling scales like $m_\phi$, the branching ratio for the higher
order decay involving scalar DM again becomes independent of the DM
mass, as in case of $\Phi \rightarrow s \phi \phi^*$ decays.


\subsection{Summary}

\begin{figure}[h]
\centering
\includegraphics[width=0.48\textwidth]{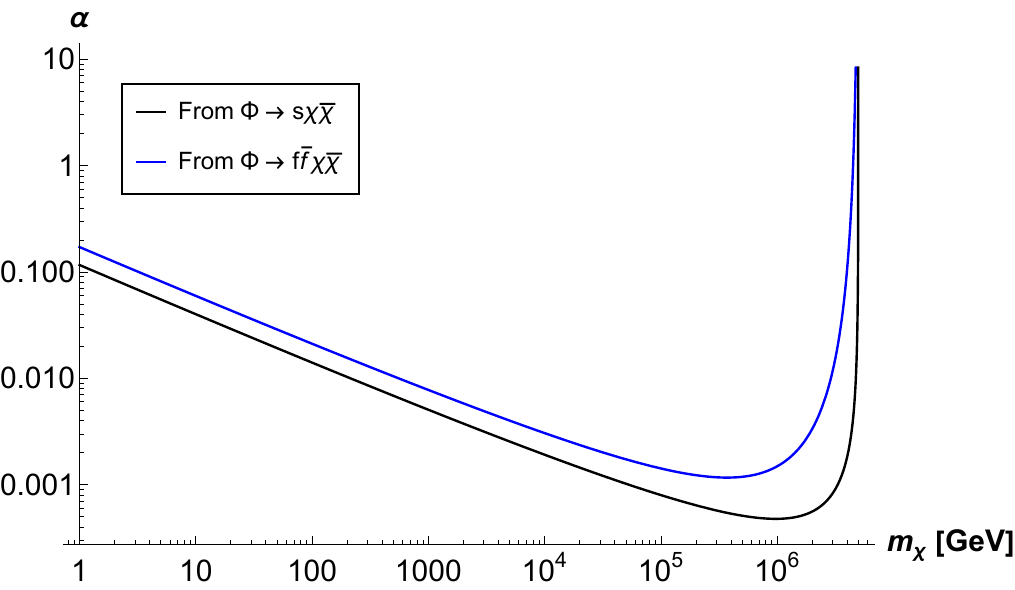} \,\,
\includegraphics[width=0.48\textwidth]{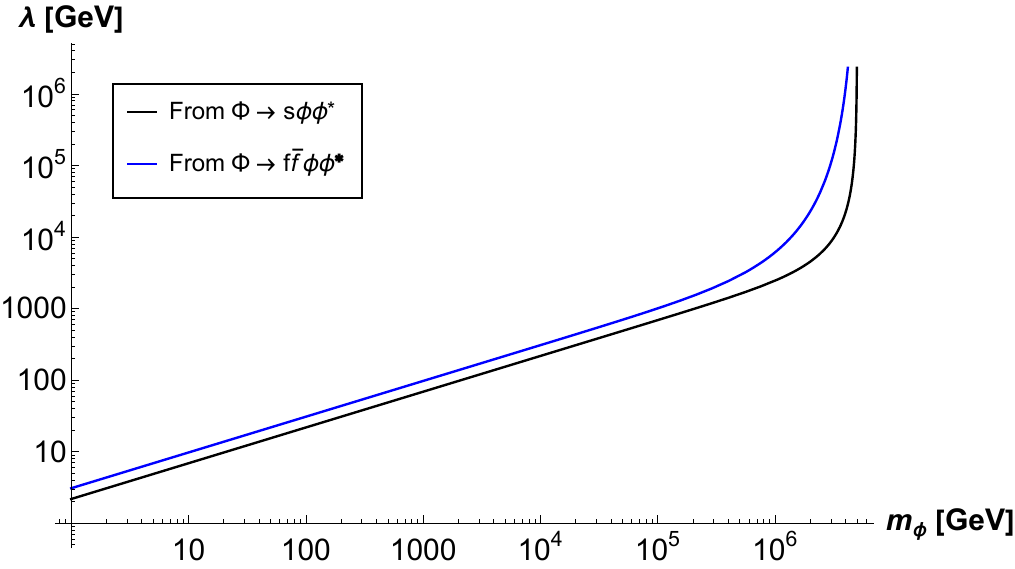} \\[5mm]
\includegraphics[width=0.48\textwidth]{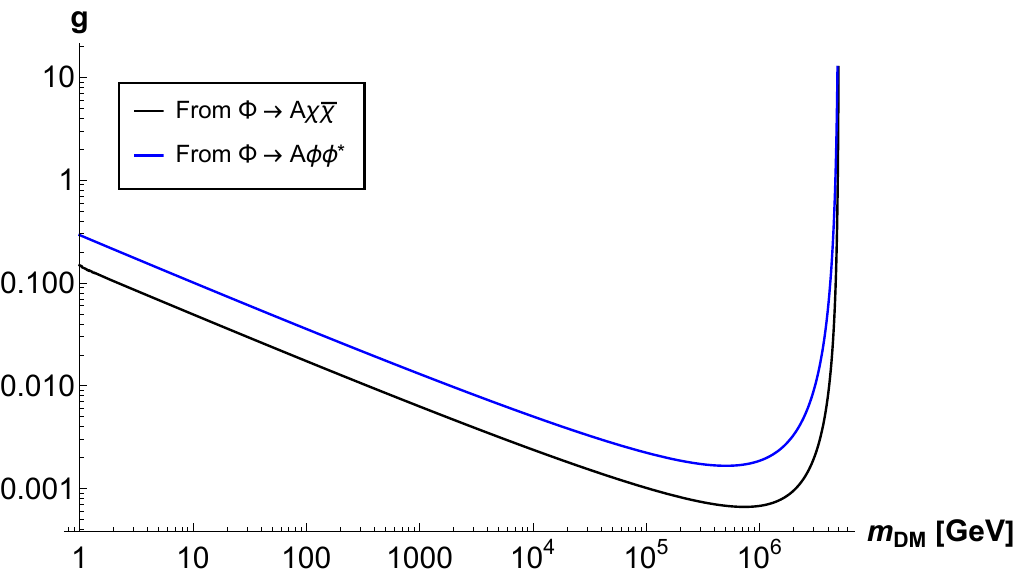}\,\,
\includegraphics[width=0.48\textwidth]{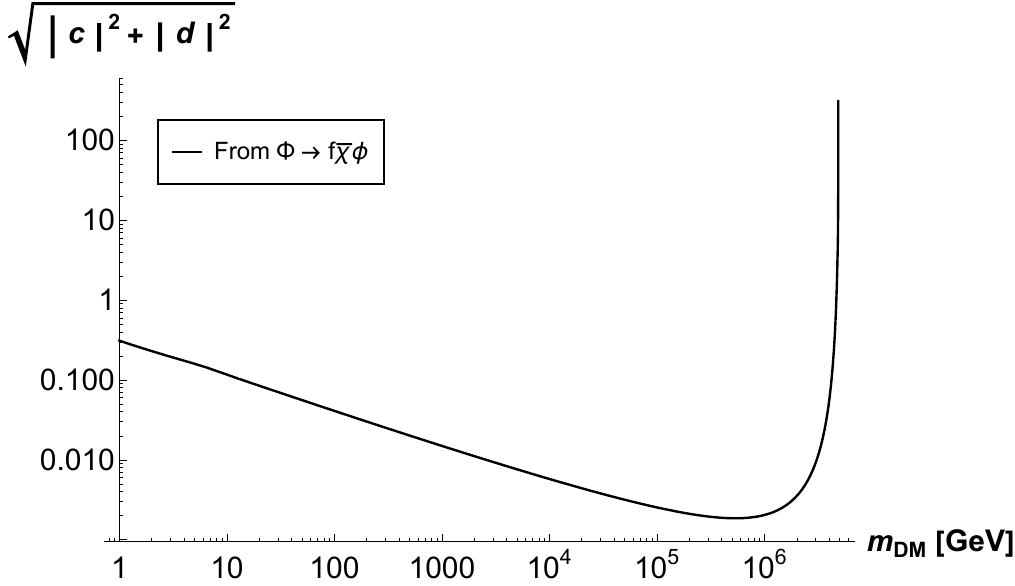}
\caption{Bounds on the coupling strengths of WIMPs derived from the
  upper bound (\ref{bbound}) on the branching ratio for $\Phi$ decays
  into WIMPs, for $m_\Phi = 10^7$ GeV. The top--left frame is for the
  $s \bar\chi \chi$ coupling $\alpha$ with $a = 0,\, b=1$
  (pseudoscalar coupling), while the top--right frame is for the
  $s \phi^* \phi$ coupling $\lambda$. In both cases, the black and
  blue curves assume that $\Phi \rightarrow s s$ and
  $\Phi \rightarrow s f \bar f$ decays dominate, respectively. The
  bottom--left frame is for the gauge coupling $g$ and fermionic
  (black) and spin--0 (blue) WIMP, under the assumption that
  $\Phi \rightarrow AA$ decays dominate. The bottom--right panel is
  for the $f \bar\chi \phi$ coupling strength $\sqrt{|c|^2+|d|^2}$,
  assuming dominant $\Phi \rightarrow s f \bar f$ decays and 
  $m_\phi = m_\chi$.}
\label{fig:summary}
\end{figure}

Having computed the branching ratios for higher--order decays into
WIMPs, in Fig.~\ref{fig:summary} we show the resulting bounds on the
WIMP couplings, derived from the WIMP mass dependent upper bound
(\ref{bbound}) on these branching ratios. In case of fermionic WIMP
$\chi$, we assume purely pseudoscalar $(a=0)$ or purely vector
$(c_A = 0)$ couplings, since this leads to slightly stronger bounds on
the overall coupling strength. Moreover, we always assume that a
single mode dominates leading order $\Phi$ decays. We observe that the
bounds on all dimensionless couplings at first strengthen with
increasing WIMP mass, scaling essentially like
$[m_{\rm DM} \ln(m_\Phi/m_{\rm DM})]^{-1/2}$, due to the explicit
$1/m_{\rm DM}$ factor in (\ref{bbound}) and the logarithmic
enhancement of the branching ratio of higher order decays mediated by
dimensionless couplings. In contrast, the bound on the dimensionful
$s \phi \phi^*$ coupling $\lambda$ (top right panel) increases
$\propto \sqrt{m_\phi}$, since the branching ratios involving this
coupling scale $\propto (\lambda/m_\phi)^2$ at small $m_\phi$.

The top--left frame shows bounds on the $s \bar\chi \chi$ coupling
$\alpha$. The bound is slightly stronger if the dominant decay mode is
$\Phi \rightarrow s s$ (black) rather than
$\Phi \rightarrow s f \bar f$ (blue), since in the former case either
of the two ``Higgs'' particles $s$ can split into a $\chi \bar\chi$
pair. Moreover, the 4--body mode $\Phi \rightarrow s \phi \bar \chi f$
suffers more relative kinematic suppression (relative to the leading
mode $\Phi \rightarrow s f \bar f$) than
$\Phi \rightarrow s \chi \bar \chi$ relative to
$\Phi \rightarrow s s$. As a result, if $\Phi \rightarrow s f \bar f$
decays dominate the bound on $\alpha$ reaches its minimum already at
$m_\chi \simeq 4 \cdot 10^5$ GeV, whereas for dominant
$\Phi \rightarrow s s$ decays the minimum is reached only at
$m_\chi \simeq 10^6$ GeV. Moreover, we see that for a ``typical'' WIMP
with mass of order 100 GeV the coupling $\alpha$ needs to be around
$0.02$ or less; this bound strengthens to values below $10^{-3}$ for
dominant $\Phi \rightarrow ss$ decays and $m_\chi \sim 10^6$
GeV. Couplings above $0.1$ are allowed only for very light
($m_\chi \lsim 1$ GeV) or very heavy ($m_\chi \gsim m_\Phi/3$) DM
particles.

Bounds on the dimensionful $s \phi \phi^*$ coupling $\lambda$ are
shown in the top right panel. We again find a stronger upper bound for
dominant $\Phi \rightarrow ss$ decays compared to dominant
$\Phi \rightarrow s f \bar f$ decays, for essentially the same reasons
as in the case of fermionic WIMP discussed in the previous paragraph. We
also note that the bound on the dimensionless ratio $\lambda / m_\phi$
is considerably weaker than that on the dimensionless couplings. For
example, $\lambda / m_\phi \geq 0.1$ is allowed for $m_\phi \lsim 500$
GeV as well as for $m_\phi \gsim m_\Phi/3$.

The bottom left panel shows bounds on the gauge coupling $g$, assuming
dominant $\Phi \rightarrow AA$ decays. For fermionic DM particle
$\chi$ the bound is quite similar to that on $\alpha$ for the case of
dominant $\Phi \rightarrow ss$ decay. For small DM mass the bound on
$g$ is two times weaker in the case of scalar DM, which contains only two
degrees of freedom, compared to four in the Dirac particle $\chi$; due
to the different threshold behaviors the ratio between the two bounds
becomes even larger close to threshold.

Finally, the bottom right panel shows bounds on the strength of the
$\phi \bar\chi f$ coupling $\sqrt{|c|^2+|d|^2}$, assuming
$m_\phi = m_\chi$. Since for the value $m_\Phi = 10^7$ GeV chosen here
one has
$\Gamma(\Phi \rightarrow s f \bar f) \gg \Gamma(\Phi \rightarrow f
\bar f)$ and
$\Gamma(\Phi \rightarrow s \phi \bar\chi f) \gg \Gamma(\Phi
\rightarrow \phi \bar\chi f)$ we ignore the decay modes without $s$
particle in the final state when computing the branching ratio. We
find slightly weaker bounds than for the other dimensionless
couplings, but nevertheless, values above $0.1$ are only allowed for
$m_{\rm DM} \lsim 10$ GeV and $m_{\rm DM} \gsim m_\Phi/3$.\footnote{We
  note that for small dark sector masses,
  $\Gamma(\Phi \rightarrow f \bar\chi \phi) / \Gamma(\Phi \rightarrow
  f \bar f) \simeq \Gamma(\Phi \rightarrow s f \bar\chi \phi) / \Gamma(\Phi
  \rightarrow s f \bar f)$; in this limit the upper bound on
  $|c|^2 + |d|^2$ from (\ref{bbound}) is therefore the same for heavy
  moduli, where $\Phi \rightarrow s f \bar f$ dominates in leading
  order, and for lighter moduli, where $\Phi \rightarrow f \bar f$
  dominates. However, due to the lower bound on $T_{\rm RH}$,
  $m_\Phi \leq 7.5$ TeV requires that the coefficient $C$ in
  eq.(\ref{phiwidth}) be larger than $1$.}

\section{Moduli Decays in SUSY}
\label{moddecayssusy}

The lightest neutralino of the minimal supersymmetric extension of the
Standard Model (MSSM) was once considered a prime example for an
attractive WIMP candidate. However, higgsino-- or wino--like
neutralinos have the correct thermal relic density in standard
cosmology for masses slightly or well above 1 TeV, respectively, in
conflict with notions of weak--scale naturalness
\cite{ArkaniHamed:2006mb, Baer:2012cf}. Moreover, indirect searches
begin to impose significant lower bounds on these WIMP candidates
independent of their production mechanism, if they account for the
bulk of DM \cite{Ahnen:2016qkx}. As the bino is a gauge singlet in the
MSSM, its mass parameter, $M_1$, cannot be constrained experimentally
if all sfermions are sufficiently heavy
\cite{Dreiner:2009ic}. However, scenarios with a light bino and heavy
sfermions lead to a very large bino relic density if the bino
decoupled after the last period of entropy production. Also, direct
searches impose very strong bounds on the amount of higgsino--bino
mixing. In standard cosmology, a bino--like neutralino can therefore
only have the right relic density if its (effective) annihilation
cross section is ``accidentally'' enhanced, e.g. via co--annihilation
with a nearly degenerate slepton \cite{Ellis:1999mm} or because its
mass is very close to half of the mass of one of the neutral Higgs
bosons of the MSSM \cite{Drees:1992am}.

This picture could change drastically in the framework of moduli
cosmology due to the dependence on additional parameters: the mass of
the modulus, its effective branching ratio into DM particles and the
reheat temperature at the end of the modulus--dominated epoch. Note
that, in contrast to simplified models, decays of the modulus into
{\em all} sparticles are relevant as the heavier sparticles will decay
into the lightest supersymmetric particle (LSP) which is the lightest
neutralino in our case. Under the assumption that sparticles do not
thermalize\footnote{The freeze--out temperature for a particle of mass
  $m$ and roughly ${\cal O}(1)$ couplings is about $m/25$, with only
  logarithmic dependence of the coupling strength. If the LSP was in
  thermal equilibrium at the end of modulus domination, when most
  moduli decay, its relic density would be similar to that in standard
  cosmology \cite{Drees:2017iod}. If the LSP is not in thermal
  equilibrium, neither are heavier sparticles.}, we may interpret the
branching ratio of the modulus to (all) sparticles as the branching
ratio to the LSP.

Here we focus on bino--like WIMPs, which satisfy current direct and
indirect DM search constraints but usually have too high a relic
density in standard cosmology as just noted. We want to find a region
in parameter space where this latter problem is solved. This requires
the reheat temperature to be below the neutralino freeze--out
temperature in standard cosmology, which imposes an upper bound on the
modulus mass for given couplings, or an upper bound on its couplings
for a given mass. Moreover, the contribution (\ref{dmdensity}) from
$\Phi \rightarrow \; {\rm DM}$ decays has to be sufficiently small. As
already noted in the Introduction, this requires the corresponding
branching ratio to be very small. We thus must arrange the couplings
of the modulus such that to leading order it (almost) only decays into
SM particles. Even in this case the branching ratio for higher order
decays involving two superparticles may well be too high. These two
issues will be investigated in detail in this section.

We begin with a discussion of possible couplings of the modulus in the
framework of supersymmetry. A gauge singlet scalar can couple to SM
particles through the following gauge invariant operators:
\begin{itemize}
\item \textbf{Higgs mode:} $\Phi H_1 \cdot H_2$, where $H_{1,2}$ are
  the two Higgs doublets of the MSSM and ``$\cdot$'' stands for the
  $SU(2)$ invariant, antisymmetric product of two
  doublets.\footnote{In principle one could also use operators
    $\Phi |H_i|^2,\; i = 1,2$; however, these do not result from a
    (softly broken) supersymmetric action.}
\item \textbf{Fermion mode:} $\Phi \overline{f_L}f_RH_i$, where
  $\overline{f_L} f_R H_i$ are the same gauge invariant operators that
  give rise to Yukawa couplings, and hence masses, of the SM fermions.
\item \textbf{Gauge mode:} $\Phi G^{a}_{\mu \nu}G^{a\,\mu \nu}$,
  $\Phi W^{i}_{\mu \nu}W^{i\,\mu \nu}$ and
  $\Phi B_{\mu \nu} B^{\mu \nu}$, where we have coupled the modulus to
  the field strength tensors of the gauge bosons in the MSSM.
\end{itemize}
The Higgs mode operator is of dimension 3 and the other two modes
involve dimension--5 operators. In fact, they are quite similar to the
couplings introduced in eq.(\ref{modSM}). However, working in a
supersymmetric framework, we need to generate these interactions from
the modification of existing SUSY Lagrangians in the MSSM, without at
the same time generating similar couplings to superparticles.

To this end, we first define a chiral, gauge singlet modulus
superfield. In superspace coordinates:
\begin{equation} \label{modsf}
\hat{\Phi} = \Phi + \sqrt{2}\theta \widetilde{\Phi} + \theta \theta F_\Phi\;,
\end{equation}
with $\widetilde{\Phi}$ being the fermionic superpartner of the
modulus $(\Phi)$ and $F_\Phi$ being the associated auxiliary
field. Note that in this framework, the scalar modulus $\Phi$ is
necessarily complex. We identify its real part as the modulus
of the previous section. The imaginary part, which behaves
as a pseudoscalar if $CP$ is conserved, needs to be studied with
care.

If the imaginary part is much heavier than the modulus proper, one
can plausibly assume that it is basically not generated in the early
universe, or else decays well before the modulus does; either way it
wouldn't affect cosmology. However, this mass difference breaks supersymmetry.
Since we are interested in modulus masses $\gsim 10^5$ GeV one would then
either have to give up on ``weak--scale'' supersymmetry, or arrange SUSY
breaking masses to be much larger for the modulus superfield than for the
MSSM sector.

Another possibility is that the pseudoscalar partner of the modulus is
an axion--like particle (ALP). This again would require large SUSY
breaking in the modulus sector, where now the ALP is much lighter than
the modulus. In such a scenario one may have to worry about modulus
decays into ALPs producing too much dark radiation. Also, since ALPs
themselves can make good Dark Matter candidates there's no need for a
(bino--like) neutralino in this case; one could choose the LSP to be
wino--like with a mass of a few hundred GeV, such that its relic density is quite small.

The perhaps most plausible assumption is that the mass splitting
between the modulus and its pseudoscalar partner is of the order of
typical soft breaking masses, and thus much smaller than $m_\Phi$. In
that case, partial widths for decays of the pseudoscalar partner into
the modulus plus two other particles (assuming it is heavier) would
almost certainly be much smaller than the decay widths of either
particle directly into SM particles. The former decays would have to
involve operators where the single modulus field is replaced by
$\Phi^2$, which are thus of higher order than the operators giving
rise to $\Phi$ decays.\footnote{Operators proportional to $|\Phi|^2$,
or $[\Phi^2 - (\Phi^{\ast})^2]$ do not contain terms
$\propto   \Re{\rm e}(\Phi)\,\Im{\rm m}(\Phi)$.} In such a scenario one effectively has two near--degenerate very massive states with very
similar lifetimes. Moreover, at least in the limit where $m_\Phi$ is
much larger than the masses of MSSM sparticles, $\Phi$ and its
pseudoscalar partner would also have almost identical branching ratios
into superparticles. In this case, decays into superparticles can be
understood via the fragmentation of (off--shell) SM particles into
sparticle pairs, which is independent of the source of the SM
particles as long as their energy is much above the sparticle
masses. In this limit, which holds for ``weak--scale'' supersymmetry,
the results we present below, assuming that only the real part
contributes, would remain unaltered. Of course, it is also possible
that for some reason the pseudoscalar partner of the modulus has a
much smaller number density than the modulus itself, in which case it
would not affect cosmology.

If soft SUSY breaking masses are much smaller than $m_\Phi$ the
``modulino'' $\widetilde{\Phi}$ will also be nearly degenerate with
the modulus field. Some production mechanisms, e.g. through the
perturbative decay of a heavy inflaton, may therefore produce
comparable numbers of moduli and modulino particles. If the dominant
couplings of $\hat{\Phi}$ respect $R$ invariance, modulino decay will
lead to final states with an odd number of sparticles, all of which
quickly decay into the LSP. Cosmological histories with similar $\Phi$
and $\widetilde{\Phi}$ densities at early times will then lead to
effective ${\cal O}(1)$ branching ratios for $\hat{\Phi} \rightarrow$
DM decays. However, unlike scalar fields, fermionic fields are usually
not excited during inflation even if their mass is much less than the
inflationary Hubble parameter, unless they have sizable Yukawa
coupling to a light scalar \cite{Prokopec:2003qd}. One can thus easily
envision scenarios where $n_\Phi \gg n_{\widetilde \Phi}$ at all
times.

In the remainder of this section, we therefore consider a purely scalar
modulus field, i.e. the real part of $\Phi$ of eq.(\ref{modsf}). In
the following subsections we construct supersymmetric actions giving
rise to the three kinds of modulus decay listed above, using the well
established rules of supersymmetry \cite{Wess:1992cp, Martin:1997ns,
  Drees:2004book}. We then compute the branching fractions for higher
order decays to sparticles, which are inevitably induced by the MSSM
Lagrangian. Whenever required, the modulus mass will be set to
$1000 \, \text{TeV}$; for $\Lambda = \overline{M}_{\rm Pl}$ this
corresponds to a reheat temperature
$T_{\text{RH}} \approx 100 \, \text{MeV}$ \cite{Drees:2017iod}. In the
final subsection, we relax the assumption $|C| = 1$ that has been used
in the bound (\ref{bbound}), showing that acceptable supersymmetric
scenarios can be constructed if $|C| \ll 1$.

\subsection{Higgs Mode}

We write the MSSM Higgs superfields as
\begin{equation}  \label{Hdoubsf}
  \hat{H}_{1,2} = H_{1,2} + \sqrt{2}\theta \widetilde{H}_{1,2}
  + \theta \theta F_{H_{1,2}} \,.
\end{equation}
The simplest supersymmetric and gauge invariant coupling between the
modulus and Higgs superfields thus is
\begin{equation} \label{modhiggssfs}
  \int d^2\theta \hat{\Phi}\hat{H_1}\cdot \hat{H_2} = F_\Phi H_1 \cdot H_2
  + F_{H_1} \cdot H_2 \Phi + F_{H_2} \cdot H_1 \Phi
  - \widetilde{\Phi} \tilde{H}_1 \cdot H_2
  - \tilde{H}_2 \cdot H_1  \widetilde{\Phi}
- \tilde{H}_1 \cdot \tilde{H}_2 \Phi \,.
\end{equation}
Since $F_\Phi$ includes a term $m_\Phi \Phi$, the first term on the
right--hand side (RHS) of eq.(\ref{modhiggssfs}) leads to bosonic
$\Phi \rightarrow H_1 H_2$ decays. However, the last term then inevitably
leads to $\Phi \rightarrow \tilde{H_1} \tilde{H_2}$ decays with the same
partial width. This coupling therefore leads to an ${\cal O}(1)$
branching ratio for $\Phi \rightarrow $ DM decays.

Instead, let us introduce SUSY breaking \cite{Luty:2005sn,
  Shirman:2009mt} through an external spurion field
\cite{Girardello:1982} $\hat{\kappa} = \theta \theta \kappa$. We may
then construct the following interaction:
\begin{equation} \label{eq:spurion}
  \mathcal{L}_{\Phi H_1 H_2} = c_H \int d^2\theta \,(\hat{\kappa} \hat{\Phi}
  \hat{H}_1 \cdot \hat{H}_2) + \text{h.c.}
  = c_H \kappa \Phi H_1 \cdot H_2 + \text{h.c.} \,.
\end{equation}
It couples the modulus only to the Higgs doublets through a soft SUSY
breaking operator, with a super--renormalizable vertex factor
$c_H \kappa$ which we refer to as $C_H$ henceforth; no (direct)
coupling of $\Phi$ to the higgsinos, or other sparticles, is
generated\footnote{A coupling of the modulus to Higgs bosons only, without coupling to higgsinos, also results from the $D-$term operator $\int d^4\theta \,\hat{\Phi} \,\hat{H_1^*} \hat{H_2^*} + \text{h.c.}$ \cite{Bae:2022okh}. The results for branching ratios into superparticles are then exactly as in the scenario investigated here.}. Note that the dimensionful quantity $C_H$ should be of order
$m^2_\Phi / M_{\rm Pl}$, and thus much smaller than typical MSSM soft
breaking masses, if the lifetime of $\Phi$ is to be similar to that
for decays due to non--renormalizable operators. Recall that the
reheat temperature immediately after $\Phi$ decays is
$\propto \sqrt{\Gamma_\Phi}$, so a shorter lifetime means higher
$T_{\rm RH}$ and correspondingly a more severe upper bound on
$B_{\Phi \rightarrow {\rm DM}}$, see eq.(\ref{dmdensity}).

We write the Higgs doublets of the MSSM as
\begin{equation} \label{eq:doublets}
    H_1 \equiv \begin{pmatrix}  
h^{0}_1 \\[1mm]
h^{-}_1
\end{pmatrix}\,, \quad 
H_2 \equiv \begin{pmatrix}  
h^{+}_2 \\[1mm]
h^{0}_2
\end{pmatrix}\,.
\end{equation}
Eq.(\ref{eq:spurion}) thus reads in components:
\begin{equation} \label{higgsia2mod}
  C_H\, \Phi \,(h_1 \cdot h_2) + \text{h.c.}
  = C_H\, \Phi (h^{0}_1 h^{0}_2 - h^{-}_1 h^{+}_2) +\, \text{h.c.}  \,.
\end{equation}
This describes the coupling of $\Phi$ to Higgs bosons in the
interaction basis. In order to compute the couplings to Higgs mass
eigenstates, we use the following relations \cite{Drees:2004book}:
\begin{subequations} \label{ia2masshiggs}
\begin{equation}
  h^\pm_1 = H^\pm \sin \beta - G^\pm \cos \beta,\ \ \ 
  h^\pm_2 = H^\pm \cos \beta + G^\pm \sin \beta \,;
\end{equation}
\begin{equation}
  \Im{\rm m} (h^0_1) = \frac{A} {\sqrt{2}} \sin \beta
  - \frac {G^0} {\sqrt{2}} \cos \beta, \ \ \
  \Im{\rm m}(h^0_2) = \frac{A} {\sqrt{2}} \cos \beta
  + \frac {G^0} {\sqrt{2}} \sin \beta \,;
\end{equation}
\begin{equation}
  \Re{\rm e}(h^0_1) = \frac {(H \cos \alpha - h \sin \alpha + v_1)} {\sqrt{2}},
  \ \ \
  \Re{\rm e}(h^0_2) = \frac {(H \sin \alpha + h \cos \alpha + v_2)} {\sqrt{2}}
  \,.
\end{equation}
\end{subequations}
Here $H$ and $h$ are the neutral CP even Higgs bosons, $A$ is the neutral
CP odd boson and $H^{\pm}$ are the charged Higgs bosons, whereas $G^0$
is the neutral Goldstone boson and $G^\pm$ are the charged Goldstone
bosons. $v_1$ and $v_2$ are VEVs acquired by the neutral Higgs fields.

As already discussed in Sec.~2, following eq.(\ref{fermiondecay}), we
will ignore mixing between $\Phi$ and the MSSM Higgs bosons but will
include decays of $\Phi$ into Goldstone modes. The relevant trilinear
couplings can be derived by substituting eq.(\ref{ia2masshiggs}) into
eq.(\ref{higgsia2mod}). Writing $\Phi = (\Phi_R + i \Phi_I)/\sqrt{2}$
with real fields $\Phi_{R,I}$ we finally find:
\begin{align} \label{modtohiggs}
  \mathcal{L}_{\Phi \mathcal{H} \mathcal{H}'} = \frac{C_H \,\Phi_R}{\sqrt{2}}
  & \bigg( \frac{1}{2} HH \sin 2\alpha - \frac{1}{2} hh \sin 2\alpha
    + Hh \cos 2\alpha - \frac{1}{2} AA  \sin 2\beta + A G^{0} \cos 2\beta
\nonumber      \\
  & + \frac {1}{2} G^0 G^0 \sin 2\beta
    -  H^+ H^- \sin 2\beta + G^-H^+ \cos 2\beta + G^+H^- \cos 2\beta
    + G^+ G^- \sin 2\beta \bigg)
\nonumber  \\
  + \frac{C_H \,\Phi_I} {\sqrt{2}}
  & \bigg( HA \cos(\alpha-\beta) - H G^0 \sin(\alpha-\beta)
    - hA \sin(\alpha-\beta) - hG^0 \cos(\alpha-\beta)
\nonumber \\
   & - i(H^- G^+ - H^+ G^-) \bigg) \,.
\end{align}
Recall that we neglect the masses of the Higgs bosons in the final state.
The total width for $\Phi_R$ decays into two MSSM Higgs bosons is then
independent of the angles $\alpha$ and $\beta$. In order to see this, note
that the coupling of $\Phi_R$ to two identical neutral bosons is two times
larger than the corresponding coefficient in the Lagrangian (\ref{modtohiggs}),
but the corresponding decay widths include an additional factor $1/2$ because
of the presence of two identical particles in the final state. The final
result is
\begin{equation} \label{GamtotHiggs}
  \Gamma_\Phi^{H} = \sum_{{\cal H, H'}} \Gamma(\Phi \rightarrow {\cal H H'})
  = \frac {|C_H|^2} {8 \pi m_\Phi}\,.
\end{equation}

Higher order decays involve diagrams where one of the Higgs or
Goldstone bosons is off--shell and couples to a pair of
electroweakinos or sfermions, as in Fig.~\ref{fig:HOPhi2h}. In other
words, we need to compute the branching ratio
\begin{equation}    \label{BRHiggs}
  \text{BR}_H =  \left\{ \sum_{\mathcal{H}} [
    \sum_{i,j} \,\Gamma(\Phi \rightarrow \mathcal{H} \tilde\chi_i \tilde\chi_j)
    + \sum_{\tilde f, \tilde f'} \,
    \Gamma(\Phi \rightarrow \mathcal{H} \tilde f \tilde f') ] \right\} / 
  \Gamma_\Phi^{H}\,.
\end{equation}
The partial widths appearing in the numerator of eq.(\ref{BRHiggs})
can be computed from generalizations of eqs.(\ref{HOScalardecay}),
given in Appendix \ref{hmodedw}; the couplings of the MSSM Higgs and
Goldstone bosons to electroweakinos and sfermions can be found in
refs.~\cite{Gunion:1989we, Drees:2004book}. In our numerical results
we take into account gaugino--higgsino mixing among the
electroweakinos as well as mixing between the components of the Higgs
doublets as described by eqs.(\ref{ia2masshiggs}), but for simplicity
neglect $\tilde f_L - \tilde f_R$ mixing, as well as possible mixing
between generations, amongst the sfermions. Note that sfermion mixing
will be numerically important only if both the relevant mixing angle
and the mass splitting between the mass eigenstates are large, which
requires some amount of finetuning.

Similarly, we consider different masses of the electroweakinos
appearing at the vertex but assume all sfermions to have a common
mass. Recall from the discussion of Figs.~\ref{fig:summary} that the
bounds on the trilinear scalar couplings are relatively weak, when
measured in units of the mass of the scalar particle in the final
state. For sfermion masses above a TeV the purely electroweak
couplings, which result from the $SU(2) \times U(1)_Y$ $D-$terms in
the scalar potential and dominate the couplings of Higgs and Goldstone
bosons to first and second generation sfermions, therefore give rise
to harmlessly small branching ratios. Unless $\tan\beta$ is very
large, the trilinear couplings $|f_\tau A_\tau|, \, |f_\tau \mu|$
will also be much smaller than the $\tilde \tau$ masses, leading to
small branching ratios into $\tilde \tau$ pairs plus a Higgs bosons;
here $f_\tau$ is the Yukawa coupling of the $\tau$ lepton, $A_\tau$ a
soft breaking parameter, and $\mu$ the supersymmetric higgsino mass
parameter. In practice the assumed common sfermion mass thus only has
to hold (at least approximately) for third generation squarks, which
can be produced with sizable branching ratios.

In particular, decays into a neutral Higgs or Goldstone boson plus a
$\tilde t \tilde t^*$ pair, or a charged Higgs or Goldstone boson plus
a $\tilde t \tilde b^*$ pair, involve the ${\cal O}(1)$ top Yukawa
coupling, and will therefore be sufficiently suppressed only if the
trilinear soft breaking parameter $|A_t|$ and $|\mu|$ are both
significantly smaller than the stop masses, or if the latter are quite
close to, or above, $m_\Phi/2$. For our numerical results we therefore
chose a common sfermion mass of $450$ TeV and $|A_t| = 50$ TeV, so
that the branching ratio into a Higgs or Goldstone boson plus
sfermions is well below $10^{-4}$.

\begin{figure}[t]
\centering
\includegraphics[width=0.8\textwidth]{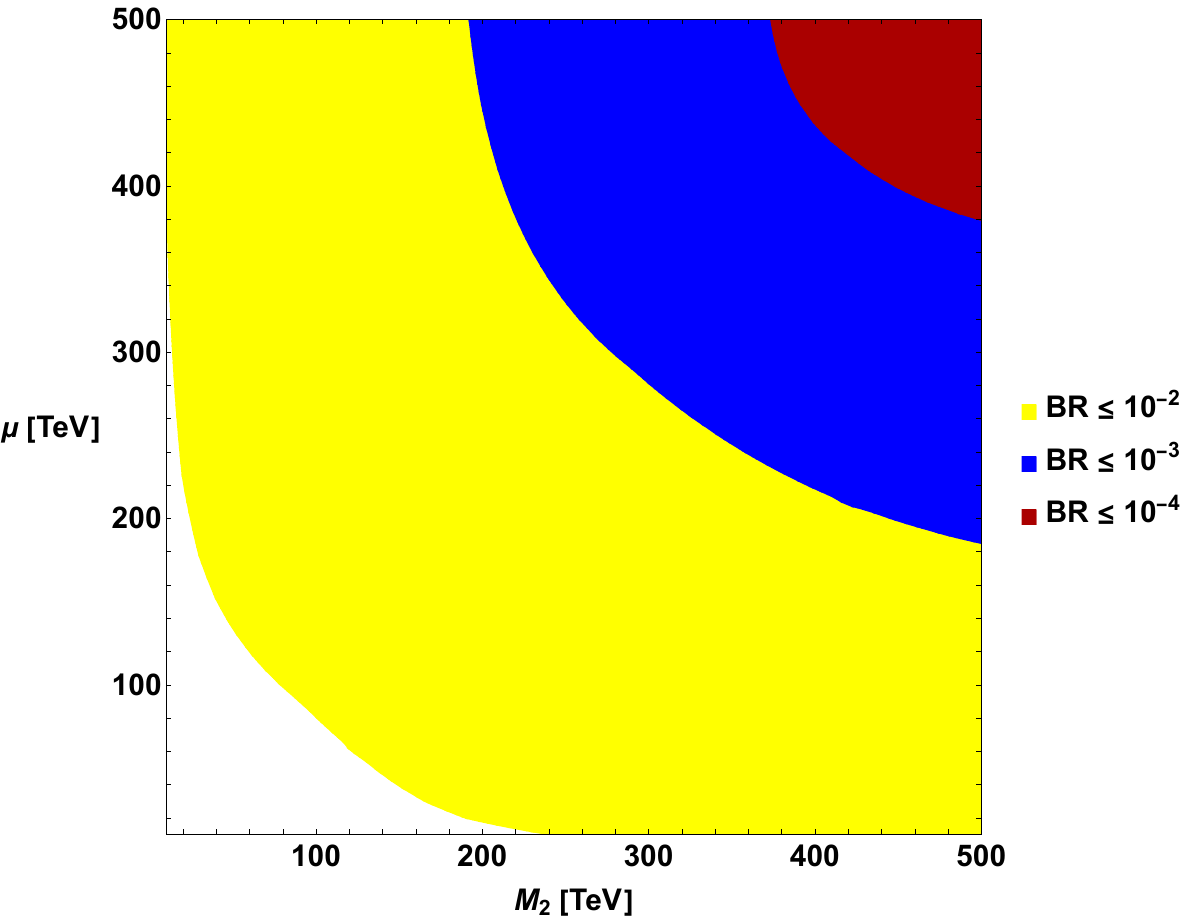}
\caption{Contours of different branching ratios from the Higgs mode in
  the plane spanned by parameters $M_2$ and $\mu$. Colors are as
  indicated in the legend. The modulus mass is taken to be 1000
  TeV. We have taken
  $\alpha_2 (m_{\Phi}) = g_2(m_\Phi)^2/(4\pi) = 0.033$. The common
  sfermion mass is taken to be 450 TeV and the trilinear coupling is
  taken to be 50 TeV.}
\label{BRH}
\end{figure}

This leaves decays into electroweakinos. The corresponding branching
ratio mostly depends on the parameters $M_1$ (bino mass), $M_2$ (wino
mass) and $\mu$.  We assume gaugino mass unification, leading to
$M_1 = \frac{M_2}{2}$. Accordingly, we perform a scan with the
remaining free parameters $M_2$ and $\mu$, presented in
Fig.~\ref{BRH}. These results are based on exact diagonalizations of
the neutralino and chargino mass matrices. We see that for electroweak
scale masses of the $\tilde \chi$ states, in fact for masses up to
about $100$ TeV, higher order processes lead to an effective branching
ratio for $\Phi$ decays into DM in excess of $1\%$. This is much above
the bound (\ref{bbound}), since our unification condition implies that
all $\tilde \chi$ masses are above 50 TeV.

The shape of the contours in Fig.~\ref{BRH} can be understood from the
observation that, in the absence of $\tilde \chi$ mixing, all MSSM Higgs
bosons couple to one higgsino and one gaugino state. Since winos
couple with $SU(2)$ gauge coupling, which is about two times larger
than the $U(1)_Y$ gauge coupling, and get additionally enhanced by a
multiplicity factor of $3$, over most of the parameter range shown the
relevant decays involve one higgsino, with mass close to $|\mu|$ and
one wino with mass close to $|M_2|$. This explains why the results are
to good approximation symmetric under the exchange
$\mu \leftrightarrow M_2$. Moreover, the branching ratio becomes very
small once $|M_2| + |\mu| \gsim m_\Phi$.  This is true even if only
one of these parameters is large. For example, $|\mu| = m_\Phi = 10^3$
TeV and $M_2 \sim 1$ TeV implies that all MSSM Higgs couplings to
$\tilde \chi$ states that are kinematically accessible in $\Phi$
decays are of order $M_Z/|\mu| \sim 10^{-4}$ at most, leading to very
small (although nonzero) $B_{\Phi \rightarrow {\rm DM}}$.  However, in
the opposite scenario, where $|\mu| \simeq 1$ TeV, both $|M_1|$ and
$|M_2|$ would need to be at least as large as $m_\Phi$ in order to
achieve $B_{\Phi \rightarrow {\rm DM}} \leq 10^{-4}$.

For most choices of the currently still allowed parameters
$M_1, \, M_2$ and $\mu$ the electroweakino masses are approximately
given by $|M_1|$ (one Bino--like two--component state), $|M_2|$ (three
Wino--like states) and $|\mu|$ (four higgsino--like states). If all these
masses are much less than $m_\Phi$, the neutralino and chargino mixing angles
will largely drop out after summing over all modes, and one obtains:
\begin{equation} \label{BRHchichi}
  B(\Phi \rightarrow H \tilde\chi \tilde\chi) \simeq \frac {\alpha_2} {4 \pi}
\left[ \frac {3}{2} \ln \frac {m_\Phi^2} {(|\mu| + |M_2|)^2} - \frac{45}{16}
  + \frac {\tan^2\theta_W} {2} \left( \ln \frac {m_\Phi^2} {(|\mu| + |M_1|)^2}
        - \frac{15}{8} \right) \right]\,,
\end{equation}
where $\alpha_2 = g_2^2/(4\pi)$ is the ``fine structure constant'' of
$SU(2)$ and $\theta_W$ the electroweak mixing angle, both taken at
scale $m_\Phi$. From Fig.~\ref{fig:schichi} we see that this
approximation should be accurate for
$|\mu|,\, |M_1|,\, |M_2| < m_\phi/10$. In particular, if all
electroweakino masses are around 1 TeV or less, eq.(\ref{BRHchichi})
predicts a branching ratio for $\Phi \rightarrow$ DM decays of at
least $5\%$.

\subsection{Fermion Mode}

We now turn to the case where the leading decay of the modulus is into
two SM fermions. We saw in the previous section that this coupling is
only gauge invariant if it, in addition, involves a Higgs field. In
supersymmetric scenarios, the relevant coupling thus arises from the
$\hat{H}\hat{f}_L\hat{\overline{f_R}}$ terms in the MSSM superpotential,
which are products of three chiral superfields. In order to couple the
gauge singlet modulus to this gauge invariant combination in SUSY, we
must consider a product of four chiral superfields:
\begin{equation} \label{quadsf}
\begin{split}
  \hat{\Phi}\hat H \hat f_L \hat{\overline {f_R}} \supset \;\Phi H
  \tilde f_L \tilde f_R^*
  + \theta \theta & (F_H \tilde f_L \tilde f_R^* \Phi
  + F_{f_L} H \tilde f_R^* \Phi
  + F_{f_R}^* H \tilde f_L \Phi
  + F_\Phi H \tilde f_L \tilde f_R^* \\
  &- \tilde h f_L \tilde f_R^* \Phi
  - f_L \overline{f_R} H \Phi
  - \overline{f_R} \tilde h \tilde f_L \Phi ) \,,
\end{split}
\end{equation}
where we have retained interaction terms involving the modulus and
wrote
$\hat{\overline{f_R}} = \tilde f_R^* + \sqrt{2} \theta \overline{f_R}
+ F_{f_R}^*$ as usual.

Such quartic terms in the superpotential are known to generate
non--renormalizable interactions \cite{Wess:1992cp, Martin:1997ns},
which is desirable to sufficiently suppress the $\Phi$ decay
width. Moreover, the penultimate term in eq.(\ref{quadsf}) contains
precisely the coupling we want. However, at the same time numerous
couplings to superparticles $\tilde f_L, \, \tilde f_R^*$ and/or
$\tilde h$ (the higgsino) are generated with identical
coefficients. These would lead to a branching ratio for $\Phi$ to DM
decays of order unity, unless all superparticles have masses above
$m_\Phi/2$. Moreover, generating the desired coupling via a spurion
superfield in this case does not work, either: it would generate the
first term on the RHS of eq.(\ref{quadsf}), which also involves
superparticles.  In fact, it does not seem to be possible to only
generate the desired $\Phi H \bar f_R f_L$ type coupling in
supersymmetric modulus scenarios. We conclude that the scenario where
the modulus dominantly decays into SM fermions (plus a Higgs boson)
cannot be realized in supersymmetry.

\subsection{Gauge Mode}

The pure gauge part of a supersymmetric Yang--Mills Lagrangian is
given by
\begin{equation}
  \mathcal{L}_{\text{gauge}} = \int d^2 \theta \, \frac {1} {4}
  \hat{W}^{aA} \hat{W}_{aA} + \text{h.c.}  \,,
\end{equation}
where $\hat{W}^A$ is the usual chiral gauge field--strength
superfield, $A$ being a 2--component spinor index and $a$ an index for
the members of the adjoint representation of the gauge
group. Multiplying this by the modulus chiral superfield $\hat{\Phi}$
generates new $F$-term contributions, including
\begin{equation} \label{gaugemod}
  \mathcal{L}_{\Phi-\text{gauge}} = - \frac {C_g} {4\sqrt{2}\Lambda} \Phi_R
  F^a_{\mu \nu} F^{a\,\mu \nu} + g^2 \frac {C_g} {2 \sqrt{2}\Lambda} \Phi_R
  (\phi^{\dag}_i T^a_{ij} \phi_j)^2 + \frac {iC_g} {2 \sqrt{2}\Lambda} \Phi_R
  \overline{\lambda}^a \slashed{\Delta}^{ac}\lambda^c\,,
\end{equation}
with $\Phi_R$ as in eq.(\ref{modtohiggs}). Here $C_g/(4\Lambda)$ is the
coefficient of the non--renormalizable
$\hat{\Phi} \hat{W}^{Aa} \hat{W}_{Aa}$ term in the superpotential.
The auxiliary fields $D^a$ have been integrated out, and we have
defined the gauge covariant derivative acting on the (4--component
Majorana) gaugino fields $\lambda^a$ as
$\Delta^{ac}_\mu = \delta^{ac} \partial_\mu + igt^{abc} A^b_\mu$.
$F^a_{\mu \nu} = \partial_\mu A^a_\nu - \partial_\nu A^a_\mu - g
t^{abc} A^b_\mu A^c_\nu$ is the usual non--Abelian gauge field
strength tensor. Finally, $T^a$ and $t^{abc}$ are the generators and
structure constants of the gauge group, and $g$ is the gauge coupling.

The first term in (\ref{gaugemod}) couples the modulus to gauge bosons
at tree level; it corresponds to the first term in eq.(\ref{modSM}).
We only require the interaction term governing the decay of the
modulus to two gauge bosons as the other terms (absent for the case of
an Abelian gauge group) are higher order in perturbation theory, being
$\mathcal{O}(g)$ or $\mathcal{O}(g^2)$.

The last two terms in eq.(\ref{gaugemod}) lead to modulus decays into
sparticles at tree level. In particular, the seconsd term couples the
modulus to four matter sfermions or Higgs bosons. However, it is of
$\mathcal{O}(g^2)$ in perturbation theory and is suppressed by
four--body phase space. This contribution should thus be much smaller
than that from the decay into a single gauge boson plus a
$\tilde f \tilde f^*$ pair, which starts at order $g$ and has only
three particles in the final state. We therefore neglect this
contribution in our numerical results. For completeness' sake, we
include the calculation for this decay width in Appendix
\ref{quintic}.

\begin{figure}[t]
\centering
\includegraphics[scale=0.6]{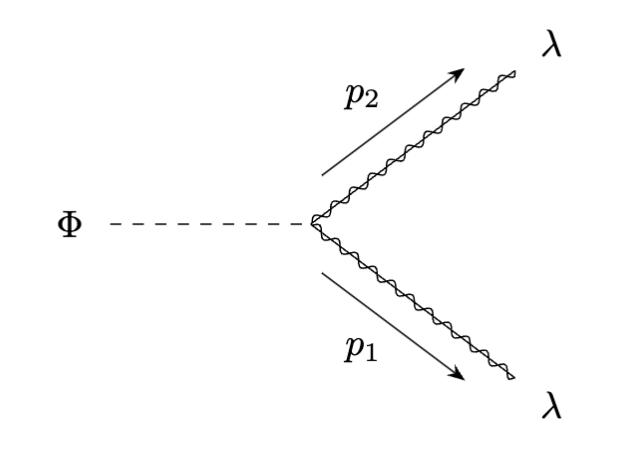}
\caption{Decay of the real part of the modulus to gauginos at tree level.}
\label{fig:modtogauginos}
\end{figure}

Expanding out the covariant derivative, we see that the third term
generates two different interaction vertices involving gauginos:
\begin{equation} \label{modgauginos}
  \mathcal{L}_{\Phi\lambda\lambda} = \frac {iC_g} {2\sqrt{2} \Lambda} \Phi
  \bar{\lambda}^a \slashed{\partial} \lambda^a
  - \frac {C_g\, gt^{abc}} {2\sqrt{2} \Lambda} \Phi \bar{\lambda}^a
  \slashed{A}^b  \lambda^{c} \,.
\end{equation}
Let us calculate the decay width for a modulus to two gauginos from
the first interaction term in (\ref{modgauginos}), see
Fig.~\ref{fig:modtogauginos}. The matrix element reads\footnote{Due to
  the Majorana nature of the gauginos there is no extra factor $1/2$
  involved when anti--symmetrizing the derivative coupling.}
\begin{align}
  i\mathcal{M} &= \frac{-C_g} {2\sqrt{2} \Lambda} \, \bar{u}(p_1) \,
  (\slashed{p_1} - \slashed{p_2})\,v(p_2) \nonumber \\
    &= \frac {-C_g} {2\sqrt{2} \Lambda} \,\bar{u}(p_1) \,(2m_\lambda)\,v(p_2) \,.
\end{align}
After squaring and summation over spins in the final state, we obtain:
\begin{align}
  \langle |\mathcal{M}|^2 \rangle  &=  \frac {C^2_g\, m^2_\lambda} {2\Lambda^2}
  \, \left[ 4 (p_2 \cdot p_1 - m^2_\lambda) \right] \nonumber \\
    &=  \frac{C^2_g\, m^2_\lambda} {\Lambda^2}\, (m^2_\Phi - 4m^2_\lambda )\;,
\end{align}
where we have used four--momentum conservation, $P = p_1 +p_2$, in order to
eliminate the dot product.

The group theory (multiplicity) factor for the gauginos is $N_g = N^2 - 1$
for an $SU(N)$ group\footnote{In the case of $U(1)$, there is only one
  gaugino so this factor is unity instead.} and we require a factor of
$\frac{1}{2}$ for indistinguishable particles in the final state. The
decay width is thus finally calculated to be \cite{Bae:2022okh}
\begin{equation}    \label{gauginodecay}
  \Gamma_{\Phi \rightarrow \lambda \overline{\lambda}} = \frac{N_g}{2}
  \int \frac {1}{2 m_\Phi} \langle |\mathcal{M}|^2 \rangle\,
  d\Pi_2 (\Phi \rightarrow \lambda \bar{\lambda})
  = N_g\, \frac {C^2_g m^3_\Phi} {64\pi \Lambda^2}
  \frac {2 m^2_\lambda} {m^2_\Phi} \left( 1 - \frac{4 m^2_\lambda } {m^2_\Phi}
  \right)^{\frac{3}{2}}\,.
\end{equation}
This decay width is proportional to the square of the gaugino mass and
will thus be very small for $m_\lambda \ll m_\Phi$. However, it
exceeds the higher order three--body contribution when
$\left( m_\lambda / m_\Phi \right)^2 > \alpha_g/\pi$; of course,
both contributions will vanish if $2 m_\lambda > m_\Phi$. Therefore
the lower bound on $m_\lambda$ required to provide sufficient phase
space suppression to gaugino modes will be set by this two--body decay
mode. Hence, we retain this decay mode in the calculation of the
branching ratio.

The second interaction term of eq.(\ref{modgauginos}), which will be
absent for the case of an Abelian gauge group, couples the modulus to
two gauginos and a gauge boson at tree level. However, this is higher
order in perturbation theory and, as we show in Appendix \ref{nonab},
also gives a contribution proportional to the gaugino mass once
diagrams are included where a gauge boson is radiated off a
gaugino. Due to these reasons, we may ignore its contribution to the
branching ratio.

\begin{figure}[t]
    \centering
\includegraphics[width=0.8\textwidth]{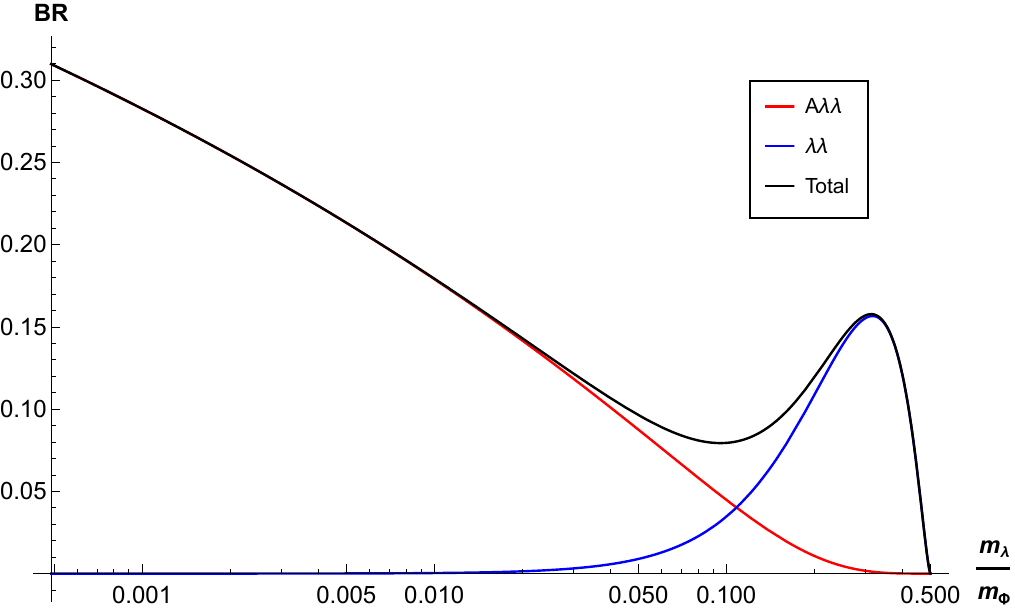}
\caption{Dependence of the branching ratios for the decay modes
  involving gauginos within an $SU(3)$ gauge theory on the gaugino
  mass, for $\alpha = \frac{g^2}{4\pi} = 0.06$. We see that the
  two--body decay to gauginos overtakes the three--body decay mode at
  $m_\lambda \simeq 0.1 m_\Phi$, and totally dominates the total
  branching ratio into gluinos for $m_\lambda \geq 0.25 m_\Phi$. Both
  channels close at the kinematic threshold $m_\lambda = m_\Phi/2$.}
\label{fig:brtogluinos}
\end{figure}

The branching ratios into final states including two gauginos, as well
as their sum, are shown in Fig.~\ref{fig:brtogluinos}, for the case of
an $SU(3)$ gauge group as in QCD. We see that the branching ratio for
the three--body mode becomes quite large for small gaugino mass
$m_\lambda$; recall that the corresponding partial width scales
$\propto \log \frac {m_\Phi} {2 m_\lambda}$, see eq.(\ref{ap3}). The
result in Fig.~\ref{fig:brtogluinos} deviates from this simple
behavior, which would lead to a linear growth for the given
logarithmic $x-$axis, since the decay widths into gauginos have also
been included in the total width, i.e. in the denominator; for
branching ratios of order $10\%$ or more this is clearly warranted. In
fact, to get an accurate result for the smallest gaugino
masses shown the logarithms should be resummed, e.g. using the
formalism of fragmentation functions that has been used some time ago
for the description of the decay of even more massive metastable
particles in the context of ultra--high energy cosmic rays
\cite{Bhattacharjee:1991zm, Barbot:2002ep, Barbot:2002gt}. However,
here we are mostly interested in identifying regions of parameter
space where the effective branching ratio into WIMPs is very small,
well below $10\%$, where resummation effects should be negligible.

\subsubsection{QCD Sector}

The interaction Lagrangian is given by
\begin{equation}
  \mathcal{L}_{\Phi-\text{QCD}} = -\frac{C_3 \delta^{ab}} {2\sqrt{2}\Lambda}
  \Phi_R [(\partial_\mu G^a_\nu) (\partial^\mu G^{b\;\nu})
  - (\partial_\mu G^a_\nu) (\partial^\nu G^{b \;\mu})]
  - \frac{i C_3 \delta^{ab}} {2\sqrt{2} \Lambda} \Phi_R \overline{\tilde{g}}^a
  \slashed{\partial} \tilde{g}^b \;,
\end{equation}
where $G$ is the gluon and $\tilde{g}$ is the gluino. The
two--body decay widths to gluons and gluinos are calculated according
to (\ref{gaugedecay}) and (\ref{gauginodecay}) with the appropriate
color factor $N_g = 8$. Sparticles can also be produced in three--body decays
involving the decay of a virtual gluon to a pair of gluinos or a pair
of squarks. Denoting the color of the gluon in the final state by $a$,
and those of the two gluinos and squarks by $b,c$ and $i,j$,
respectively, the associated color factors are:
\begin{subequations}
\begin{equation}  \label{cfggluinos}
  CF_{\Phi \rightarrow g\tilde{g}\tilde{g}} = \sum_{a,b,c,d,e} \delta^{ad} \delta^{ae}
  f^{bdc} f^{bec} = \sum_{a,b,c} f^{bac}f^{bac}
  = 2(3)^2 \frac{(3^2 -1)}{2\times3} =  24 \;,
\end{equation}
\begin{equation}    \label{cfgsquarks}
  CF_{\Phi \rightarrow g\tilde{q}\tilde{q}^{*}} = \sum_{a,d,e} \sum_{i,j} \delta^{ad}
  \delta^{ae} T^d_{ij} T^e_{ji} = \sum_a \text{tr}[T^a T^a]
  =  \sum_a \frac{\delta^{aa}}{2} = 4 \;.
\end{equation}
\end{subequations}
The partial decay widths for these modes are obtained from
eqs.(\ref{HOgaugedecay}) considering a pure vector--like
interaction. In the numerical results we assume a common squark
mass. The branching ratio to sparticles, and hence to Dark Matter, is
thus given by
\begin{equation}
  B_{\Phi \rightarrow {\rm sparticles}}^{\text{QCD}} = \frac {24\;
    \Gamma_{\Phi \rightarrow A \chi \bar\chi}
  + 4\; \sum_{\tilde q}\Gamma_{\Phi \rightarrow A \phi \phi^*}
  + \Gamma_{\Phi \rightarrow \lambda \bar\lambda} }
{\Gamma_\Phi^{\rm tot}};
\end{equation}
note that the color factor for $\Phi$ decays into gauginos is already
included in eq.(\ref{gauginodecay}). As in Fig.~\ref{fig:brtogluinos},
we have included the supersymmetric decay modes in the denominator.

\begin{figure}[t!]
\centering
\includegraphics[width=0.8\textwidth]{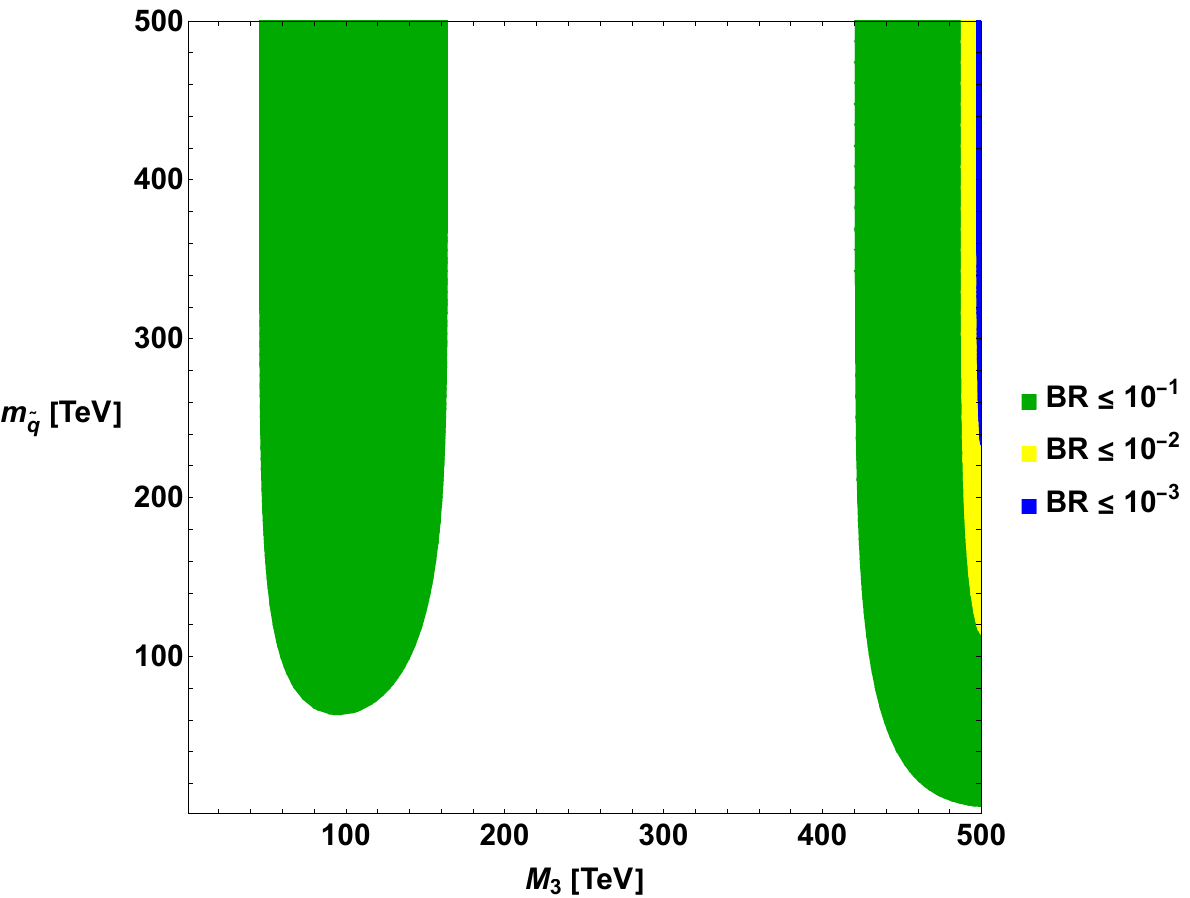}
\caption{Contours of different branching ratios into superparticles
  for the $SU(3)_C$ gauge mode in the plane spanned by the gluino mass
  $M_3$ and the common squark mass $m_{\tilde q}$. The colors are as
  indicated in the legend; in the white region the branching ratio
  exceeds $10^{-1}$. We have used $\alpha_s (m_\Phi) = 0.06$.}
\label{BRC}
\end{figure}

This allows us to perform a scan for the gluino mass ($M_3$) and the
common squark mass $(m_{\tilde q})$. The results are shown in
Fig.~\ref{BRC}. We find that a branching ratio of order $10^{-4}$ can
only be obtained at the very edge of phase space, when $M_3$ or
$m_{\tilde q} \approx \frac{m_\Phi}{2}$, the other mass being at least
as large (not shown). This can easily be inferred from the fact that a
branching ratio below $10^{-3}$ is found only in a very thin strip (in
blue) at the edge of the figure. In fact, for an extended region of
parameter space (shown in white) the total branching ratio into
superparticles even exceeds $10^{-1}$. In the region $180 \ {\rm TeV}
\leq M_3 \leq 420$ TeV this is mostly due to $\Phi \rightarrow \tilde g
\tilde g$ two--body decays, see Fig.~\ref{fig:brtogluinos}. Note that
$\Gamma_{\Phi \rightarrow \lambda \lambda} / \Gamma_{\Phi \rightarrow A A}$
is independent of the gauge group, depending only on the gaugino
mass $m_\lambda$. The same range of gaugino masses will therefore lead
to branching ratios into sparticles above $10\%$ also for dominant decays
into $U(1)_Y$ or $SU(2)$ gauge bosons.

In the limit where the gluino and squark masses are much smaller than
the modulus mass, we can directly use the approximate results of the
higher order gauge decays given in eqs.(\ref{apgauge}) in
Sec.~\ref{moddecays}, with a purely vector--like coupling (i.e.
$c_A = 0$). We can neglect the decay width of the two--body decay to
gluinos in this limit. Assuming 12 degenerate squarks, we find:
\begin{equation} \label{eq:brc}
  r^C = \frac {\Gamma_{\Phi \rightarrow {\rm sparticles}}}
  {\Gamma_{\Phi \rightarrow gg}} \simeq \frac{\alpha_s}{\pi}\left[
  2 \log \left( \frac {m^2_\Phi} {4m^2_{\tilde g}} \right)
  + \log \left( \frac{m^2_\Phi } {4m^2_{\tilde q}} \right)
  - \frac{29}{4} \right]\,.
\end{equation}
In terms of this ratio, the effective branching ratio into Dark Matter
is given by $B_{\Phi \rightarrow {\rm DM}} = \frac {2 r^C} {1 + r^C}$;
the factor of $2$ occurs because all decays of $\Phi$ into sparticles
produces two Dark Matter particles. We saw in Figs.~\ref{fig:Achichi}
and \ref{fig:Aphiphi} that the analytical approximation used in
eq.(\ref{eq:brc}) is very accurate for
$m_{\tilde g, \tilde q} \leq 0.03 m_\Phi$, which includes all cases of
``weak--scale'' supersymmetry for typical modulus masses. However, in
the region of applicability of eq.(\ref{eq:brc}) the effective
branching ratio into DM particles is clearly several orders of
magnitude above the bound (\ref{bbound}). For example, if all
sparticle masses are $\leq 0.01 m_\Phi$, which can be considered a
very loose definition of ``weak--scale'' supersymmetry for
$m_\Phi \geq 500$ TeV, we find $r^C \geq 0.31$, i.e.
$B_{\Phi \rightarrow {\rm DM}} \geq 0.47$; here we have taken
$\alpha_S = 0.06$ as in Fig.~\ref{BRC}.

\subsubsection{Electroweak Sector}

Since modulus couplings to gauge bosons are diagonal in the
interaction basis, it is more convenient to stay within this
basis. The relevant gauge bosons are thus $B_\mu$ for $U(1)_Y$, and
$W^3_\mu$ and $W^\pm_\mu$ for $SU(2)$; their fermionic superpartners
are $\widetilde{B},\, \widetilde{W}^3$ and $\widetilde{W}^\pm$. The
relevant part of the Lagrangian is given by
\begin{equation}
\begin{split}
  \mathcal{L}_{\Phi-\text{EW}} = -\frac {\Phi_R} {4\sqrt{2} \Lambda}
  &\left(
    C_Y B^{\mu \nu} B_{\mu \nu}
    + 2i C_Y \overline{\widetilde{B}} \slashed{\partial} \widetilde{B}
  + C_2 W^{3\mu \nu} W^3_{\mu \nu}
  + 2i C_2 \overline{\widetilde{W}^{3}}\slashed{\partial} \widetilde{W}^{3}
  \right. \\ & \left.
 + 2C_2 W^{+\mu \nu}W^{-}_{\mu \nu}
 + 4 i C_2\overline{\widetilde{W}^{+}} \slashed{\partial} \widetilde{W}^{-}
 \right)\;.
\end{split}
\end{equation}
The two--body decays of the modulus to a pair of electroweak gauge
bosons or a pair of binos or winos is again calculated using the
previously established expressions (\ref{gaugedecay}) and
(\ref{gauginodecay}) with relevant group theory factors. Higher--order
decays involve one of the gauge bosons in the final state being
off--shell and transitioning into a pair of electroweakinos or
sfermions. In order to derive the corresponding partial widths, we
first derive the couplings of the $B$ and $W^3$ gauge fields to
electroweakinos and sfermions, after which we modify the decay widths
accordingly and list the results in Appendix \ref{ewmodedw}. The
branching ratios are hence given by
\begin{subequations}
\begin{equation}     \label{BRY}
  B^Y_{\Phi \rightarrow {\rm sparticles}} = \frac {\sum_{\tilde\chi}
  \Gamma_{\Phi\rightarrow B \tilde\chi\tilde\chi}
  + \sum_{\tilde f} \Gamma_{\Phi \rightarrow B \tilde{f} \tilde{f}^{*}}
  + \Gamma_{\Phi \rightarrow \widetilde{B}\widetilde{B}}} {\Gamma_\Phi^{\rm tot}} \;,
\end{equation}
\begin{equation}    \label{BRW}
  B^W_{\Phi \rightarrow {\rm sparticles}} = \sum_{W^3,\;W^\pm}
  \frac { \sum_{\tilde\chi} \Gamma_{\Phi \rightarrow W \tilde\chi \tilde\chi}
    + \sum_{\tilde f} \Gamma_{\Phi \rightarrow W \tilde{f} \tilde{f}^{*}}
    + 3\;\Gamma_{\Phi \rightarrow \widetilde{W}\widetilde{W}}}
  {\Gamma_\Phi^{\rm tot}}\;.
\end{equation}
\end{subequations}
Since $U(1)_Y$ is an Abelian group, gauginos do not contribute to the
three--body decays in eq.(\ref{BRY}), but the higgsinos do, along with
all sfermions of the MSSM, since they all carry hypercharge. Assuming
relatively small $M_1$, so that
$\Phi \rightarrow \widetilde{B} \widetilde{B}$ two--body decays are
negligible, we can choose the higgsino mass parameter $\mu$ and a
common sfermion mass $m_{\tilde f}$ as free
parameters.\footnote{Assuming gaugino mass unification a small $M_1$
  implies a small $M_2 = 2 M_1$ as well, hence there are two light
  neutralinos and one light chargino in the spectrum. However, the
  $U(1)_Y$ gauge boson only couples to the higgsino components of
  these light states. The three--body partial widths into these light
  states are therefore suppressed by a factor of order $(M_Z/\mu)^4$
  if both $\tilde\chi$ are light, or ${\cal O}(M_Z/\mu)^2$ for the
  associate production of a light $\tilde\chi$ and a heavier
  higgsino--like state, making them completely negligible for
  $|\mu| > 10$ TeV.}

\begin{figure}[t]
\centering
\includegraphics[width=0.8\textwidth]{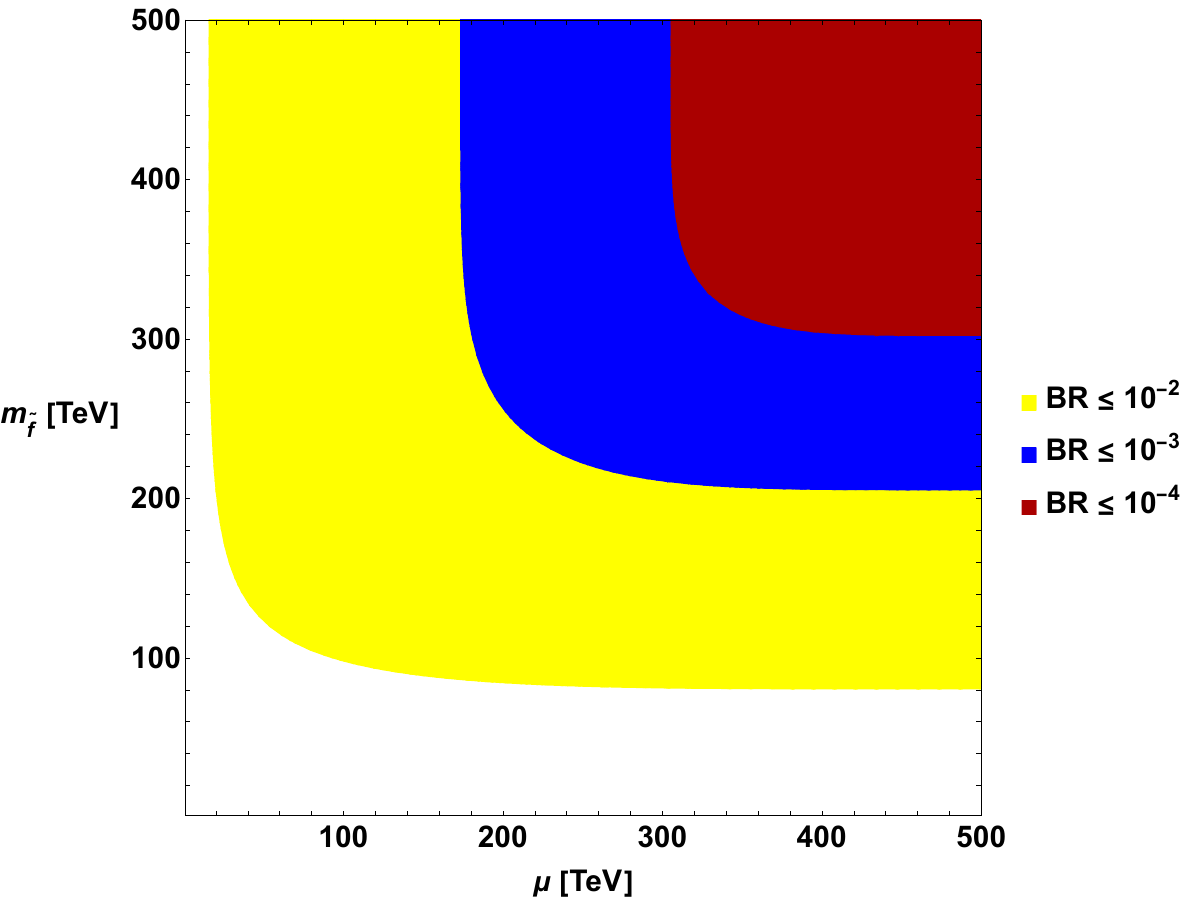}
\caption{Contours of different branching ratios from the $U(1)_Y$
  gauge mode in the plane spanned by the common sfermion mass
  $m_{\tilde f}$ and higgsino mass $\mu$. We have used
  $\alpha_Y (m_\Phi) = 0.02$. Colors are as indicated in the
  legend. We have set $M_1 = 500$ GeV so that the partial width for
  $\Phi \rightarrow \widetilde{B} \widetilde{B}$ is very small; the
  value of $M_2$ plays essentially no role here, since the hypercharge
  gauge boson does not couple to $SU(2)$ gauginos.}
\label{fig:BRY}
\end{figure}

In Fig.~\ref{fig:BRY} we show results for the total branching ratio into
superparticles assuming the modulus field only couples to hypercharge
gauge bosons and gauginos, in the plane spanned by $\mu$ and $m_{\tilde f}$.
We see that the branching ratio exceeds $1\%$ for $m_{\tilde f} \leq 100$ TeV
or $\mu < 20$ TeV; a branching ratio below $10^{-4}$ can be achieved only
if both parameters are larger than $300$ TeV.

If all sparticles are much lighter than the modulus we can again use the
approximate expressions of eqs.(\ref{apgauge}). As for the Higgs mode, we
neglect gaugino--higgsino as well as $\tilde f_L - \tilde f_R$ mixing;
this should be a good approximation as far as the total branching ratio
into superparticles is concerned. We then find:
\begin{equation} \label{eq:rY}
  r^Y = \frac{\Gamma_{\Phi \rightarrow {\rm sparticles}}}
  {\Gamma_{\Phi \rightarrow BB}} = \frac{ \alpha_Y} {2\pi}
  \bigg[ \frac {2}{3} \ln \left( \frac {m^2_\Phi} {4 |\mu|^2} \right)
  + \frac{11}{6} \ln \left( \frac {m^2_\Phi} {4 m^2_{\tilde q}} \right)
  + \frac{1}{2} \ln \left( \frac {m^2_\Phi} {4 m^2_{\tilde l_L}} \right)
  + \ln \left( \frac {m^2_\Phi} {4 m^2_{\tilde l_R}} \right)
  - \frac{149}{12} \bigg]\,.
\end{equation}
Here we have assumed a common squark mass $m_{\tilde q}$, a common
mass for all $SU(2)$ doublet sleptons $m_{\tilde l_L}$ and a third
mass for all $SU(2)$ singlet sleptons $m_{\tilde l_R}$. As before the
effective branching ratio into Dark Matter particles
$B_{\Phi \rightarrow {\rm DM}} = 2 r^Y / (1+r^Y)$. Evidently, even for
loosely defined ``weak--scale'' supersymmetry the effective branching
ratio will be ${\cal O}(10\%)$. For example, assuming all sparticle
masses to be below $0.01 m_\Phi$ again, we find $r^Y \geq 0.06$, i.e.
$B_{\Phi \rightarrow {\rm DM}} \geq 0.11$; here we have taken
$\alpha_Y = 0.02$ as in Fig.~\ref{fig:BRY}.

\begin{figure}[t]
\centering
\includegraphics[scale=0.46]{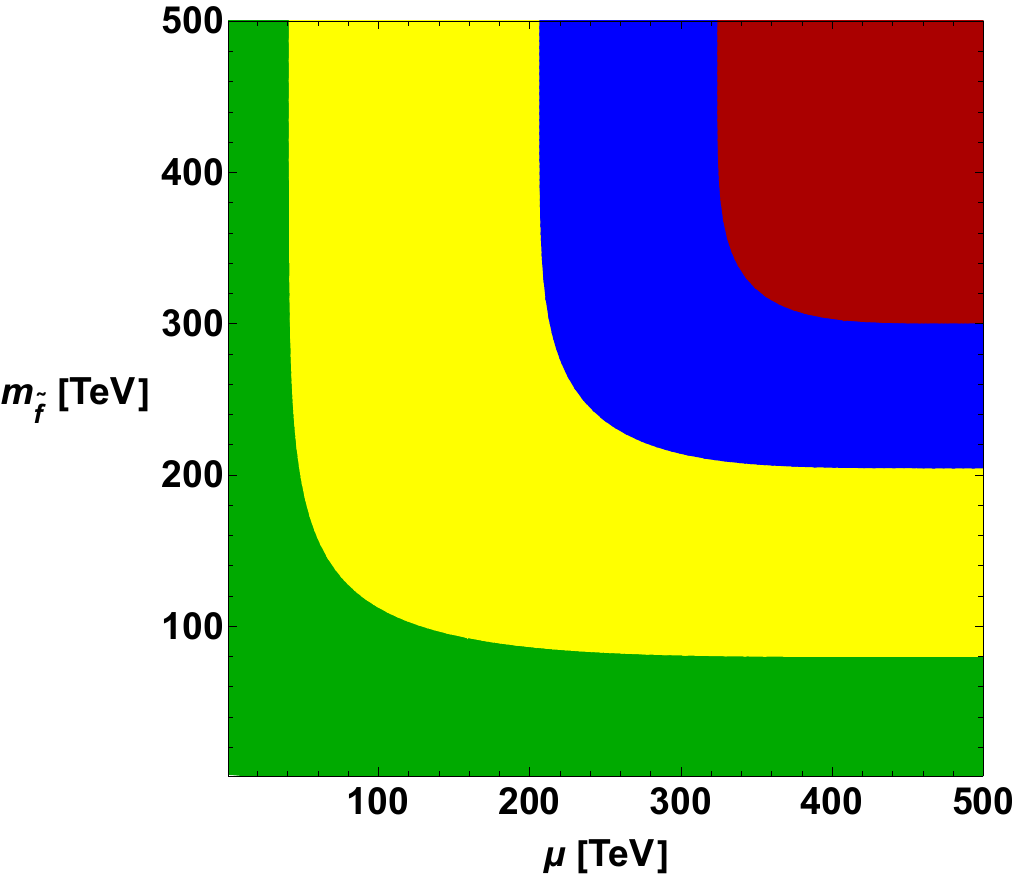}\,\,
\includegraphics[scale=0.46]{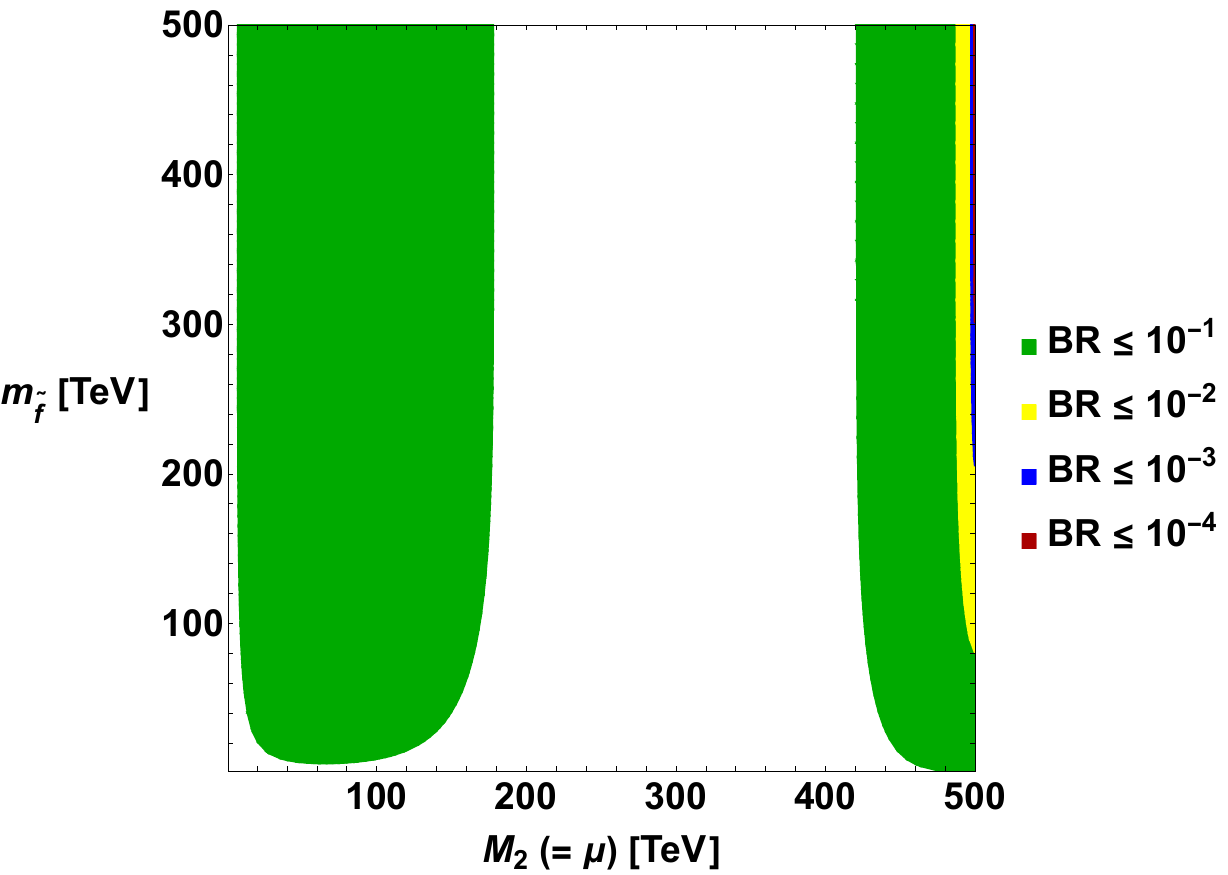}
\caption{Contours of different branching ratios from the $SU(2)$
  gauge mode. The left frame shows results in the
  $(m_{\tilde f}, \ \mu)$ plane for $M_2 > m_\Phi/2$, while the right
  frame is for $M_2 = \mu$. We have used $\alpha_2 (m_\Phi) = 0.033$
  and $M_1 = 1$ TeV. Colors are as indicated in the legend.}
\label{fig:BRW}
\end{figure}

We now turn to the $SU(2)$ case, where the gauge bosons do couple to
$SU(2)$ gauginos, as well as to higgsinos, but not to the bino. The
relevant parameters are hence $M_2$, $\mu$ and the sfermion masses. In
our numerical work, we again assume all sfermions to have a common mass
$m_{\tilde f}$. In the left frame of Fig.~\ref{fig:BRW} we take
$M_2 > m_\Phi/2$, so that gauginos cannot be produced. The results
resemble those in Fig.~\ref{fig:BRY}, where decays into gauginos also
play no role; however, the branching ratio is somewhat larger in the
$SU(2)$ case due to the larger gauge coupling. In the right frame, we
instead assume $M_2 = \mu$.  The result then resembles that for
QCD. In particular, the large white region centered at a gaugino mass
near $300$ TeV is the same in both cases, while the smaller gauge
coupling somewhat reduces the branching ratio into superparticles for
smaller masses.

In the limit of small sparticle masses, the relevant three--body widths
can again be computed from eqs.(\ref{apgauge}). We find:
\begin{align} \label{eq:rW}
r^W &= \frac {\Gamma_{\Phi \rightarrow {\rm sparticles}}}
  {\Gamma_{\Phi \rightarrow WW}} \nonumber \\
  &= \frac {\alpha_2}{\pi} \left[
    \frac {4}{3} \ln \left( \frac {m_\Phi^2} {4 |M_2|^2} \right)
    + \frac {1}{3} \ln \left( \frac {m_\Phi^2} {4 |\mu|^2} \right)
    +\frac {3}{4} \ln \left( \frac {m_\Phi^2} {4 m_{\tilde q_L}^2} \right)
    +\frac {1}{4} \ln \left( \frac {m_\Phi^2} {4 m_{\tilde l_L}^2} \right)
    - \frac {475} {72} \right]\,.
\end{align}
Again adopting a very loose definition of weak--scale supersymmetry,
the requirement that all sparticle masses should be
$\leq 0.01 m_\Phi$, we find $r^W \geq 0.16$, i.e.
$B_{\Phi \rightarrow {\rm DM}} \geq 0.27$.

\subsection{Summary}

In this section, we investigated scenarios in which the lightest
superparticle (LSP) is our WIMP candidate. In this case, not only the
direct decay of $\Phi$ particles to LSPs contributes to the DM
density, but {\em all} $\Phi$ decays into superparticles, which will
quickly decay into the LSP. Even if two--body decays of $\Phi$
primarily produce SM particles, which is true for the Higgs and gauge
modes, there will be at least some superparticles with roughly
${\cal O}(1)$ couplings to the primary $\Phi$ decay products. As a
result, for weak--scale supersymmetry the effective
$B_{\Phi \rightarrow {\rm DM}}$ is typically $\gsim 10\,\%$ \footnote{Decays of a scalar modulus field into superparticles have also been analysed in ref.\cite{Baer:2023bbn}. However, only two--body decays are included there, hence the branching ratio into superparticles is greatly underestimated if the modulus mass
lies well above the sparticle mass scale, taken to be a few TeV in that work.}. The bound
(\ref{bbound}) can only be saturated at the very edge of phase space,
where (some) superparticles have mass just under $m_\Phi/2$. Producing
the correct amount of Dark Matter from $\Phi$ decay would then require
quite extreme finetuning.

\begin{figure}[t]
\centering
\includegraphics[width=0.9\textwidth]{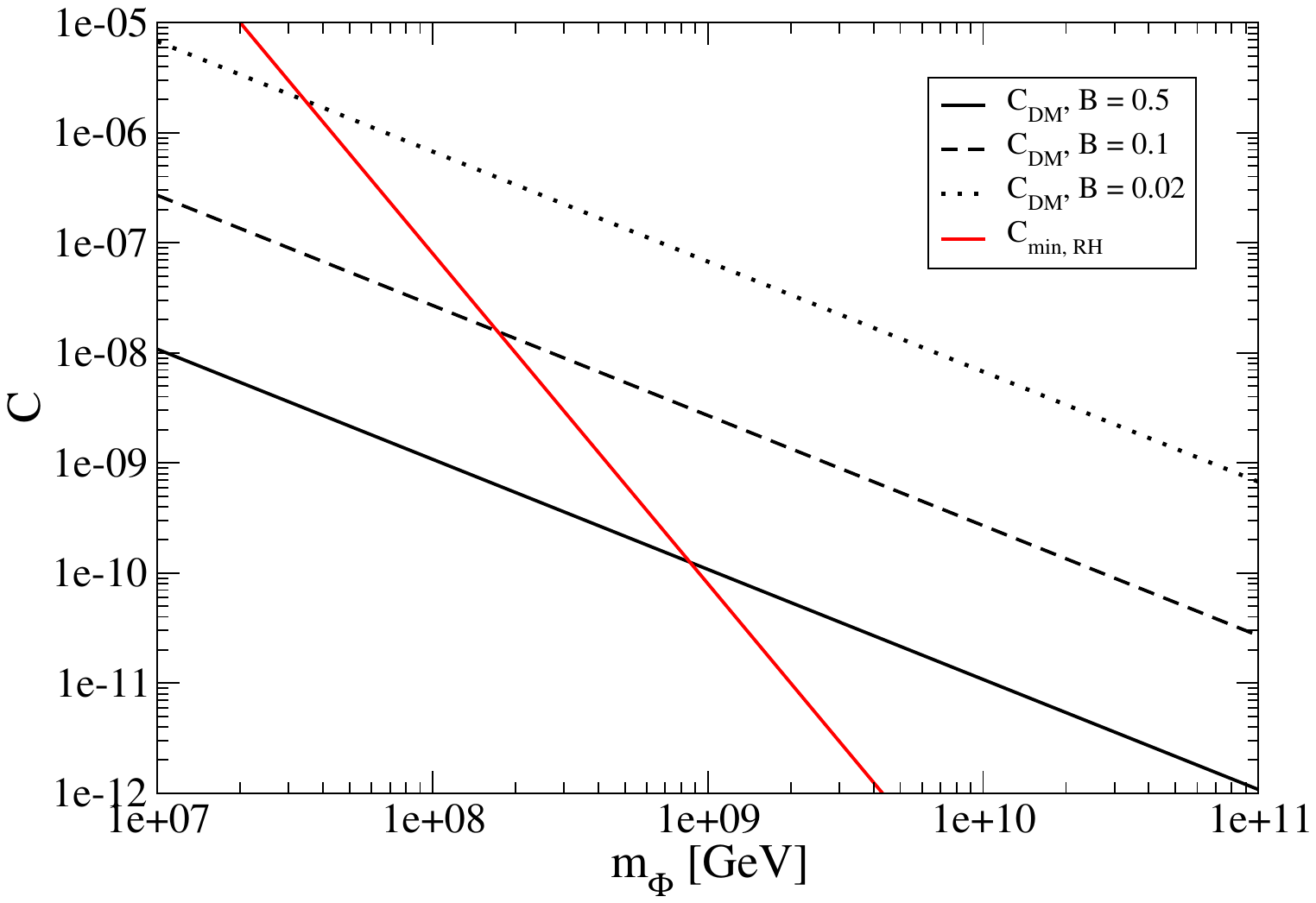}
\caption{The black lines show the value of the dimensionless modulus
  coupling $C$ that is required to obtain the desired DM relic
  density, $\Omega_{\rm DM} h^2 = 0.12$, for $m_{\rm DM} = 100$ GeV and
  $B_{\Phi \rightarrow {\rm DM}}= 0.5$ (solid), $0.1$ (dashed) or
  $0.02$ (dotted); these have been computed from eq.(\ref{C_DM}), which
  assumes that annihilations of superparticles after $\Phi$ decay, as well
  as thermal production of superparticles, are negligible. The red line
  shows the lower bound on $C$ from the requirement $T_{\rm RH} \geq 5$
  MeV, see eq.(\ref{C_RH}).}
\label{fig:C}
\end{figure}

Recall, however, that (\ref{bbound}) assumed the coefficient $C$
introduced in eq.(\ref{phiwidth}) to be $1$. This establishes a
one--to--one relation between the $\Phi$ mass and width, where the
latter determines the reheat temperature via \cite{Kolb:1990vq}
\begin{equation} \label{TRH}
  T_{\rm RH} = \left[ \frac {45} {4 \pi^3 g_*(T_{\rm RH})} \right]^{1/4}
  \sqrt{\Gamma_\Phi M_{\rm Pl}} \simeq 0.3 \sqrt{\Gamma_\Phi M_{\rm Pl}}\,.
\end{equation}
Here $\Gamma_\Phi$ is the total $\Phi$ decay width and $g_*$ is the
effective number of relativistic degrees of freedom (defined via the
energy density). Inserting eq.(\ref{phiwidth}) into eq.(\ref{TRH}),
and then inserting the result into eq.(\ref{dmdensity}) which holds
for general $C$ as long as DM annihilation after $\Phi$ decay is
negligible, we find that $\Phi$ decays produce the correct DM density
if
\begin{equation} \label{C_DM}
  C \simeq \frac {30 \ {\rm GeV}} {m_\Phi}
  \left( \frac {0.1} {B_{\Phi \rightarrow {\rm DM}}} \frac {30 \ {\rm GeV}}
    {m_{\rm DM}} \right)^2\,.
\end{equation}
This is shown as the black lines in Fig.~\ref{fig:C}, for a typical WIMP
mass of $100$ GeV and three characteristic values of $B_{\Phi \rightarrow
  {\rm DM}}$.

At the same time, successful BBN requires $T_{\rm RH} \gsim 5$ MeV
\cite{deSalas:2015glj}. Using eqs.(\ref{phiwidth}) and (\ref{TRH})
this implies
\begin{equation} \label{C_RH}
  C \geq 0.08 \left( \frac {10^6 \ {\rm GeV}} {m_\Phi} \right)^3\,.
\end{equation}
This lower bound is shown by the red line in Fig.~\ref{fig:C}. This
lower bound is compatible with the DM requirement (\ref{C_DM}) only if
\begin{equation} \label{mphimin}
  m_\Phi \geq 1.7\cdot 10^6 \,m_{\rm DM} \, \left(
    \frac {B_{\Phi \rightarrow {\rm DM}}}  {0.1} \right)\,,
\end{equation}
and
\begin{equation} \label{Cmax}
  C \leq 6 \cdot 10^{-7} \left( \frac {0.1} {B_{\Phi \rightarrow {\rm DM}}}
    \frac {30 \, {\rm GeV}} {m_{\rm DM}} \right)^3\,.
\end{equation}

We conclude that any scenario where (some) sparticles are much lighter
than $m_\Phi$, in particular weak--scale supersymmetry, requires
either a modulus with large mass but very small coupling, $|C| \ll 1$;
or a very light LSP, which would need to be bino--like in order to
evade experimental constraints \cite{Dreiner:2009ic},
$|M_1| \lsim 0.3$ GeV. Otherwise, the reheat temperature would have to
be sufficiently large, so that superparticles thermalize after $\Phi$
decay. The latter case, however, corresponds to the usual scenario of
thermal WIMPs.

\section{Conclusions}\label{conclude}

In this work, we investigated the production of Dark Matter (DM)
particles in the decay of a scalar particle $\Phi$ (modulus), under
the assumption that it is sufficiently long--lived to dominate the
energy density of the universe leading to an early matter--dominated
epoch. In this cosmological scenario the dependence of the
present--day DM abundance on additional parameters, such as the
modulus mass and branching ratio of moduli decays to WIMPs, can be
used to increase or decrease their relic abundance compared to the
baseline scenario of a thermal WIMP in standard cosmology. In
particular, if WIMPs do not thermalize after modulus decay and modulus
decay proceeds through couplings of order $1/M_{\rm Pl}$, the
effective branching ratio for $\Phi \rightarrow {\rm DM}$ decay must
be quite small, of order $10^{-4}$ for a WIMP mass of $100$ GeV (and
even less for heavier WIMPs), see eq.(\ref{bbound}). Branching ratios
of this order or larger can easily be achieved even in higher--order
decays; they have not been studied systematically before, and are
therefore the focus of this work.

First, we considered simplified models with various leading order
modulus decay modes into SM particles. One noticeable result is that,
due to gauge invariance, decays into SM fermions can proceed only via
the VEV of the Higgs field $S$ that breaks the electroweak symmetry,
or in three--body $S f \bar f$ final states, the latter being dominant
for $m_\Phi > 7.5$ TeV; in that case higher--order decays will have
(at least) four particles in the final state.

We asserted that moduli decay to WIMPs must occur at higher order if
there is any interaction between the WIMP and SM particles. We
analyzed both scalar and spin$-1/2$ WIMPs, and considered scalar
(Higgs) or spin$-1$ particles as mediators. As shown in
Fig.~\ref{fig:summary}, the requirement (\ref{bbound}) leads to upper
bounds of order (a few times) $10^{-2}$ on the strengths of
dimensionless mediator couplings and WIMP mass of 100 GeV; the
dimensionful coupling of a scalar WIMP to a scalar mediator must be
about ten times smaller than the WIMP mass. In the framework of
simplified models, this can quite easily be arranged; we note that this
region of parameter space would typically lead to (much) too large
relic densities for our WIMP candidates in standard cosmology. Of course,
the reduced couplings of the mediators will reduce the signal in both direct
and indirect WIMP searches. ${\cal O}(1)$ couplings of the mediators are
allowed only if the WIMP mass is very close to $m_\Phi/2$, which does not
seem to be very plausible.

Having established the general machinery, we next turned to the case
of the MSSM, with the lightest neutralino as WIMP candidate. Here, we
first had to check whether gauge invariant couplings could be
generated from a SUSY Lagrangian, such that the modulus primarily
decays into SM particles, with superparticles being produced only via
suppressed couplings or in higher--order diagrams.  We found that the
modulus could couple to the Higgs sector of the MSSM via soft breaking
of SUSY and to the separate gauge sectors of the MSSM in a SUSY
invariant fashion. The latter ``gauge mode'' coupling also introduces a
tree--level decay of the modulus to gauginos, but the corresponding
matrix element is proportional to the gaugino mass. However, it is not
possible to couple a gauge singlet modulus to only SM--like fermions
at tree--level in (softly broken) supersymmetry; hence we discarded
the ``fermion mode'' in our numerical analysis of the MSSM.

It is important to note that in the MSSM $\Phi$ decays into {\em any}
superparticle will contribute to the effective branching ratio for
$\Phi \rightarrow {\rm DM}$ decays, since they will all (quite
quickly) decay into the LSP (plus SM particles).  We therefore
analyzed the total branching ratio of $\Phi$ into superparticles in
case of leading decay into Higgs bosons or SM gauge bosons. Since
there are always some superparticles with roughly ${\cal O}(1)$
couplings to any given SM particle, we found that the bound
(\ref{bbound}) can only be saturated, i.e. just the right amount of DM
can only be produced in $\Phi$ decays in supersymmetry, if these
superparticles have masses very, very close to $m_\Phi/2$, which
requires severe finetuning.  On the other hand, we saw in
Fig.~\ref{fig:C} that the correct relic density can be obtained
without much finetuning if the modulus coupling to the SM sector is of
order $10^{-3}/M_{\rm Pl}$ or less; the lower bound on the reheat
temperature then leads to a lower bound on $m_\Phi$ of order $10^7$
GeV or more.  These constraints pose a severe challenge to realistic
models of moduli (or similar particles). Alternatively one can
consider a very light, bino--like LSP, which can only be realized if
gaugino masses do not unify.

As noted, in our analysis we assumed that the long--lived particle is
a scalar (CP--even spin$-0$).  One can also envision matter domination
by other particles, e.g. long--lived fermions \cite{Allahverdi:2022zqr}.
In detail, this will require a dedicated, new analysis, but we expect the
main conclusions to carry over also to these cases.

\appendix
\numberwithin{equation}{section}

\section{Phase Space Factorization} \label{phasespacefac}

The Lorentz Invariant Phase Space (LIPS) for a final state with $n$
particles, each carrying four--momentum $p_i$, is given by
\begin{equation} \label{LIPS}
  d\Pi_{\text{LIPS}} =  \left( \prod_{i\,=\,1}^n \frac {1} {(2\pi)^3}
    \frac {d^3\vec{p_i}} {2E_i} \right) \, (2\pi)^4 \,
  \delta^{(4)}(P - \sum_{i\,=\,1}^n p_i) \;.
\end{equation}
Here $P$ is the sum of the four--momenta of the initial states. For a
single decaying particle, $P$ is simply its four--momentum; in
particular $P = (m_\Phi, \vec{0})$ for us, since we wish to calculate decay
widths in the rest frame of the decaying particle. The phase space may
be partitioned or ``factorized'' by collecting the first $j$ particles in the
final state into the subsystem $X$ and the remaining $n-j$ particles
into the subsystem $Y$ \cite{Barger:1987book}:
\begin{align}
  d\Pi_{\text{LIPS}} &=  \frac {d m^2_X} {2\pi}\, \frac {d m^2_Y} {2\pi} \,
  \frac {d^3 \vec{p_X}} {(2\pi)^3 2E_X} \frac {d^3 \vec{p_Y}} {(2\pi)^3 2E_Y} \,
  (2\pi)^4 \delta^{(4)}(p_Z - p_X - p_Y) \nonumber \\
  & \quad
  \cdot \left( \prod_{i\,=\,1}^j \frac {1} {(2\pi)^3} \frac {d^3\vec{p_i}} {2E_i}
  \right) \, (2\pi)^4 \delta^{(4)}(p_X -\sum_{i\,=\,1}^jp_i)\, \left(
    \prod_{k\,=\,j+1}^n \frac {1} {(2\pi)^3} \frac {d^3\vec{p_k}} {2E_k} \right)
  (2\pi)^4 \delta^{(4)}(p_Y -\sum_{k\,=\,j+1}^{n}p_k) \nonumber \\
  &= \frac {d m^2_X} {2\pi} \, \frac {d m^2_Y} {2\pi} \,
  d\Pi(Z \rightarrow XY) \, d\Pi(X \rightarrow (1,2,3,...j))\,
  d\Pi(Y \rightarrow (j+1,j+2,...n))\,.
\end{align}
The limits on the ``masses'' $m_X$ and $m_Y$ are
\begin{equation}
  \sum^j_{i=1} m_i \leq m_X\,, \quad \sum^n_{k=j+1} m_k \leq m_Y,\quad
  m_X +m_Y \leq m_Z \;.
\end{equation}
Each subsystem is by itself Lorentz invariant and hence all
calculations can be carried out in individual, convenient frames of
reference.

For reference, we quote the analytical result for a two--body phase space:
\begin{equation}   \label{2bodyres}
  \int d\Pi_{\text{2}} = \frac {\uplambda^{\frac{1}{2}}(M^2,{m_1}^2,{m_2}^2)}
  {8 \pi M^2} \;,
\end{equation}
where $\uplambda(x,y,z) = x^2 + y^2 + z^2 -2xy -2yz - 2zx$ is the
K\"{a}ll\'{e}n function.

In this work, we mainly require the factorization of three- and four--body
phase spaces. A three--body phase space can be factorized
into ``one--body'' and two--body sub-systems:
\begin{equation} \label{3bodyfac}
  d\Pi(Z \rightarrow 123) = \frac {d m^2_X}{2\pi}\, d\Pi(Z \rightarrow 1X) \,
  d\Pi(X \rightarrow 23)\,.
\end{equation}
A four--body phase space can be factorized in two different ways:
by breaking into either two two--body sub-systems, or into
``one--body'' and three-body sub--systems:
\begin{subequations}
\begin{equation}  \label{4body22}
  d\Pi(Z \rightarrow 1234) = \frac {d m^2_X}{2\pi} \, \frac{d m^2_Y} {2\pi}
  d\Pi(Z \rightarrow XY) \, d\Pi(X \rightarrow 12)\, d\Pi(Y \rightarrow 34) \;;
\end{equation}
\begin{equation}  \label{4body31}
  d\Pi(Z \rightarrow 1234) = \frac {d m^2_X} {2\pi}\, d\Pi(Z \rightarrow 1X) \,
  d\Pi(X \rightarrow 234) \;.
\end{equation}
\end{subequations}
In (\ref{3bodyfac}) and (\ref{4body22}), one can rewrite the matrix
element squared accordingly and use (\ref{2bodyres}). For
(\ref{4body31}), the three--body subsystem can be dealt with through
use of scaled variables by using the ``mass'' of the subsystem as the
scaling.

\section{SUSY Decay Widths}

\subsection{Quintic Scalar Coupling}\label{quintic}

\begin{figure}[H]
\centering
\includegraphics[scale=1]{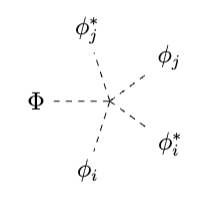}
 \caption{Quintic scalar coupling leading to the decay of the modulus
   $\Phi$ to four MSSM scalars, $i$ and $j$ being ``flavor'' labels.
   For non--Abelian gauge groups the $\phi \phi^*$ pairs should be in
   the adjoint of the gauge group, see eq.(\ref{gaugemod}).}
\label{fig:quinscalar}
\end{figure}

Here, we treat the decay of the modulus to four MSSM scalars via the
$D^2$ term, as depicted in Fig. \ref{fig:quinscalar}. 

No propagators are involved so this is essentially a matter of
computing the group and multiplicity factors and integrating purely
over the phase space. We use (\ref{4body22}) for this purpose, and
note that the matrix element is a constant.
\begin{align}   \label{quinscalarres}
  \Gamma_{\Phi \rightarrow \phi_i \phi^*_i \phi_j \phi*_j} &=
  \frac{\langle |\mathcal{M}|^2 \rangle} {2 m_\Phi} \int \frac{dm^2_X}{2\pi}\,
  \frac{dm^2_Y}{2\pi}\, \int d\Pi_2(\Phi \rightarrow XY) \,d
  \Pi_2(X \rightarrow \phi_i \phi^*_i) \,d\Pi_2(Y \rightarrow \phi_j \phi^*_j) \nonumber 
  \\
  &= \frac{C^2_g}{16\pi} \, \frac{\alpha^2_g}{32\pi^2}\, \frac{N_{ij} \kappa_g}
  {\Lambda^2 \, m^3_{\Phi}} \,\int \int
  \uplambda^{\frac{1}{2}}(m^2_\Phi, m^2_X, m^2_Y) \,
  \sqrt{1 - \frac {4 m^2_{\phi_i}} {m^2_X} }\,
  \sqrt{1 - \frac{4 m^2_{\phi_j}}{m^2_Y} } dm^2_X dm^2_Y\;,
\end{align}
where $C_g$ is the coupling factor from eq.(\ref{gaugemod}). The group
factor $\kappa_g = \sum_{a,b} [ \tr (T^a T^b) ]^2$ for non--Abelian
interactions evaluates to $N/4$ for an $SU(N)$ group; for $U(1)_Y$
interactions, $\kappa_1 = (Y_i Y_j)^2$. $N_{ij}$ is a multiplicity
factor (e.g. $N_{ij} = 9$ for two $\tilde q \tilde q^*$ pairs produced
via electroweak interactions, and $N_{ij} = 4$ for two
$\tilde q_L \tilde q_L^*$ pairs produced via $SU(3)$
interactions). Finally, the integration limits are
\begin{equation}
    4m^2_{\widetilde{q_j}} \leq m^2_Y \leq (m_{\Phi} - m_X)^2, \quad
    4m^2_{\widetilde{q_i}}\leq m^2_X \leq (m_{\Phi} - m_{\widetilde{q_j}})^2\,.
    \notag
\end{equation}

From (\ref{quinscalarres}), we can ascertain that this channel is
firstly, higher order in perturbation theory, indicated by the presence
of an extra $\alpha_g^2$ factor and secondly, this decay mode is also
suppressed by phase space. The resulting branching ratio will
therefore be far smaller than that for
$\Phi \rightarrow A \phi_i \phi_i^*$ three--body decays, and can
safely be neglected.

\subsection{Non-Abelian Vertex: Modulus--Gauginos--Gauge}
\label{nonab}

\begin{figure}[t]
\centering
\includegraphics[scale=0.6]{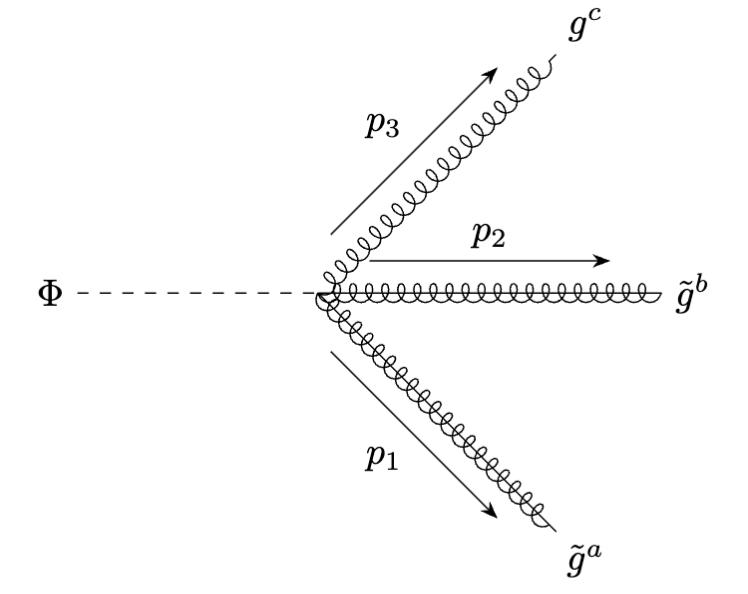}
\caption{Tree--level decay of the modulus $\Phi$ to gluinos and a gluon due to
   the non-Abelian vertex.}
\label{fig:modgauginosgauge}
\end{figure}

Due to the non--Abelian nature of the $SU(2)$ and $SU(3)_C$ gauge
groups, there is an additional vertex generated from the covariant
derivative coupling the modulus to two gauginos and a gauge
boson.\footnote{There is also a similar derivative coupling to three
  gauge bosons, but this can be considered a higher--order correction
  to the leading decay into two gauge bosons, and will therefore be
  ignored.} This diagram is problematic as it is of the same order as
the higher order decays of the modulus to gauginos, and would result
in interference due to the presence of the same final states. We will
now examine this vertex in detail, by looking at the case of $SU(3)_C$
in Fig. \ref{fig:modgauginosgauge}, since here this contribution would
be most dangerous due to the presence of the strong interaction
coupling. In particular, we will show that this extra contribution to
the matrix element also scales like the gluino mass for
$m_{\tilde g} \ll m_\Phi$, once all relevant diagrams are included, as
implied by the equation of motion. The couplings of $SU(2)$ can be
treated in an analogous manner.

\begin{figure} [b]
\centering
\includegraphics[scale=0.8]{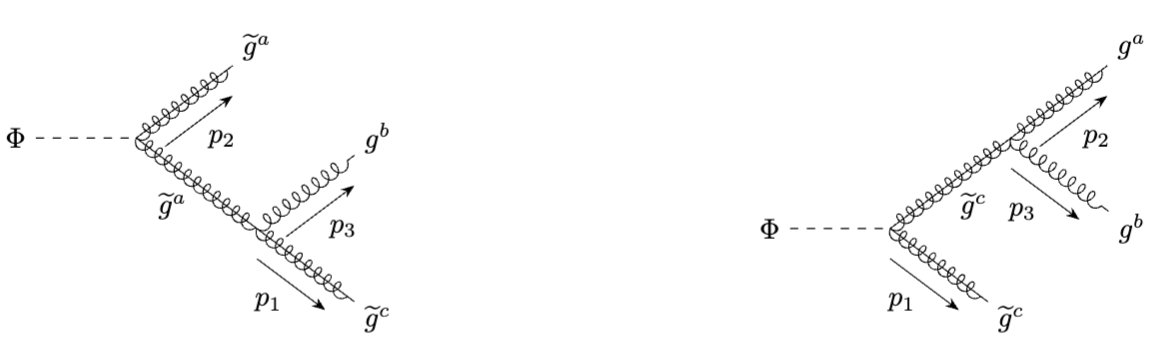}
\caption{Two possible diagrams for radiation of a gluon off of a gluino.}
  \label{fig:nonabinter}
\end{figure}

The relevant matrix element is
\begin{equation}    \label{modgaugegaugino}
  i\mathcal{M} = -i \frac{C_3 g_s}{\sqrt{2}\Lambda} t^{abc} \left[
    \bar{u}(p_1) \gamma^{\mu} v(p_2) \right]\, \epsilon^{*}_{\mu}(p_3) \;.
\end{equation}
Before we calculate the decay rate, we take a step back and consider
if there exist other diagrams of the same order that may
contribute. In particular, the diagrams in Fig. \ref{fig:nonabinter}
evidently lead to the same final state.

The corresponding matrix elements are, with $p_3$ being the momentum
of the gluon:
\begin{subequations}
\begin{align}
  i\mathcal{M}_1 &= (-g_s t^{abc})\bar{u}(p_1) \gamma^{\mu} \left[
    \frac{i((\slashed{p_1}+\slashed{p_3}) + m_{\tilde g})}{(p_1 + p_3)^2
      - m^2_{\tilde g}} \right] \frac{-C_3}{2\sqrt{2}\Lambda}\,\left[
    (\slashed{p_1}+\slashed{p_3}) - \slashed{p_2} \right]\,v(p_2) \,
  \epsilon^{*}_{\mu}(p_3) \nonumber
    \\
    &= \frac{i C_3 g_s t^{abc}} {2\sqrt{2}\Lambda}\, \bar{u}(p_1) \left[
      \frac{\gamma^{\mu}((\slashed{p_1}+\slashed{p_3}) + m_{\tilde g})
        ((\slashed{p_1}+\slashed{p_3}) + m_{\tilde g})}{(p_1 + p_3)^2
        - m^2_{\tilde g}}\right] v(p_2) \epsilon^{*}_{\mu}(p_3) \;,
\end{align}

\begin{align}
  i\mathcal{M}_2 &= \bar{u}(p_1)\frac{-C_3}{2\sqrt{2}\Lambda} \left[
    (\slashed{p_1}+\slashed{p_3}) - \slashed{p_2} \right]
  \left[\frac{-i((\slashed{p_2}+\slashed{p_3}) - m_{\tilde g})}{(p_2 + p_3)^2
      - m^2_{\tilde g}} \right] (-g_s  t^{abc}) \gamma^{\mu} v(p_2)
  \epsilon^{*}_{\mu}(p_3) \nonumber \\
  &= \frac{i C_3 g_s t^{abc}}{2\sqrt{2}\Lambda} \bar{u}(p_1) \left[
    \frac{((\slashed{p_2}+\slashed{p_3}) + m_{\tilde g})
      ((\slashed{p_2}+\slashed{p_3}) + m_{\tilde g})\gamma^{\mu}}{(p_2 + p_3)^2
      - m^2_{\tilde g}}\right]v(p_2) \epsilon^{*}_{\mu}(p_3) \;.
\end{align}
\end{subequations}
In the limit of vanishing gluino mass, we find
\begin{align}
  i\mathcal{M}_1 + i\mathcal{M}_2 &\stackrel{(m_{\tilde g} \rightarrow 0)}
  = \frac{i C_3g_s t^{abc}}{2\sqrt{2}\Lambda}\bar{u}(p_1) \left[
    \frac{\gamma^{\mu} (\slashed{p_1}+\slashed{p_3})
      (\slashed{p_1}+\slashed{p_3})}{(p_1 + p_g)^2}
    + \frac{(\slashed{p_2}+\slashed{p_3}) (\slashed{p_2}+\slashed{p_3})
      \gamma^{\mu}}{(p_2 + p_3)^2} \right] v(p_2)\,\epsilon^{*}_{\mu}(p_3) \nonumber \\
  &= \frac{i C_3 g_s t^{abc}}{2\sqrt{2}\Lambda} \bar{u}(p_1) \left[
    \frac{\gamma^{\mu} (p_1 + p_3)^2}{(p_1 + p_3)^2} + \frac{(p_2 + p_3)^2
      \gamma^{\mu}}{(p_2 + p_3)^2} \right] v(p_2)\epsilon^{*}_{\mu}(p_3) \nonumber \\
  &= \frac{i C_3 g_s}{\sqrt{2}\Lambda} t^{abc} \left[ \bar{u}(p_1) \gamma^{\mu}
    v(p_2) \right] \epsilon^{*}_{\mu}(p_3) \;,
\end{align}
which exactly cancels the matrix element (\ref{modgaugegaugino}) from
Fig. \ref{fig:modgauginosgauge}. We hence conclude that the
four--point coupling of the modulus to two gauginos and a gauge boson
can be considered a higher order in correction to the leading
three--point coupling to two gauginos, both leading to contributions
that are proportional to the gaugino mass. For these reasons, we drop
the contribution of this decay mode in our branching ratio
calculations.

\subsection{Higgs Mode} \label{hmodedw}

The higher order decay widths for a final state containing an MSSM
Higgs or Goldstone boson and a pair or electroweakinos or sfermions
are given respectively by:
\begin{subequations}
\begin{align}
  \Gamma_{\Phi \rightarrow \mathcal{H} \tilde\chi_i\tilde\chi_j}
  = \frac {C^2_H} {16\pi m_\Phi} \frac {\alpha_2 \text{S}^2_{\mathcal{H}}} {4\pi}
 & \int_{(m_{\tilde\chi_i} + m_{\tilde\chi_j})^2}^{m^2_\Phi}
 \bigg\{ \frac{ (|A_{ij}|^2 + |B_{ij}|^2) (m^2_X - m^2_{\tilde\chi_i}
    - m^2_{\tilde\chi_j}) - 2 (|A_{ij}|^2 -|B_{ij}|^2) m_{\tilde\chi_i}
    m_{\tilde\chi_j}} {m^4_X} \nonumber \\
  & \cdot \left( 1 - \frac {m^2_X} {m^2_\Phi} \right)
    \frac{\uplambda^{\frac{1}{2}} (m^2_X, m^2_{\tilde\chi_i}, m^2_{\tilde\chi_j})}
    {m^2_X} \bigg\} \, dm^2_X \;.
\end{align}
\begin{equation}
  \Gamma_{\Phi \rightarrow  \mathcal{H} \tilde{f}\tilde{f'}}
  = \frac {C^2_H} {16 \pi m_\Phi} \frac{\alpha_2 |\Delta_{st}|^2
    \text{S}^2_{\mathcal{H}}} {8\pi m^2_\Phi} \int_{4 m^2_{\tilde{f}}}^{m^2_\Phi}
  \frac {m^2_\Phi} {m^4_X} \left( 1 - \frac {m^2_X} {m^2_\Phi} \right)
  \sqrt{ 1 - \frac{4 m^2_{\tilde{f}}} {m^2_X} }  dm^2_X \;.
\end{equation}
\end{subequations}
Here $\alpha_2 = \frac{g^2_2}{4 \pi}$ is the fine structure constant of
$SU(2)$, and $\text{S}_{\mathcal{H}}$ is
a constant containing all relevant symmetry factors and mixing angles
for the relevant MSSM Higgs bosons. $A_{ij}$, $B_{ij}$ are linear
combinations of mixing matrices of the electroweakinos and
$\Delta_{st}$ is a dimensionful parameter describing the coupling of the
Higgs bosons to sfermions, which depends on $\tilde f_L - \tilde f_R$
mixing in the sfermion sector.

\subsection{Electroweak Gauge Mode} \label{ewmodedw}

Here, we list partial widths for the higher order decays from the
$U(1)_Y$ sector. Decays from the $SU(2)$ sector are treated similarly.

For the higher order decays to electroweakinos,
\begin{align}     \label{dwBelwinos}
  \Gamma_{\Phi \rightarrow B \tilde\chi_i \tilde\chi_j}
  = \frac{C^2_Y m^3_\Phi} {32\pi \Lambda^2} \frac {\alpha_Y} {4\pi}
  \int \int
  &\bigg\{ \frac {1} {(1-x_3)^2}
  \bigg[ 2 (|C_{ij}|^2 - |D_{ij}|^2)
    (\sqrt{\mu_i \mu_j} x^2_3) \nonumber \\
  & + (|C_{ij}|^2 + |D_{ij}|^2) \big[x_3 ( (1-x_3 + \mu_j - \mu_i)
       (1 - x_1 + \mu_j - \mu_{i}) \nonumber\\
  &\hspace*{5mm} - (1 - x_3 + \mu_i - \mu_j) ( 1 - x_1 - x_3 + \mu_i- \mu_j))
  \\
  &\hspace*{5mm} + 2(1-x_3) ( 1 - x_1 + \mu_j - \mu_i)
    (1 - x_1 - x_3 + \mu_j - \mu_i)\big] \bigg]\bigg\}  \, \,dx_3 \,dx_1 \;.
\nonumber
\end{align}
Here, $C_{ij}$ and $D_{ij}$ are mixing matrix elements of the
electroweakinos, and we have defined
$\mu_k = m^2_{\tilde\chi_k}/m^2_\Phi$. The limits of integration are:
\begin{equation*}
  x_3 \lessgtr \frac{ (2 - x_1) (1 + \mu_j - \mu_i - x_1) \pm
    \sqrt{ (x^2_1 - 4\mu_j)}\, (1 + \mu_j - \mu_i - x_1)} {2\,(1-x_1+\mu_j)}\;,
\end{equation*}
\begin{equation*}
  2\sqrt{\mu_j} <\, x_1 < 1 + \mu_j - \mu_i\;.
\end{equation*}

For the higher order decays to sfermions,
\begin{equation}    \label{dwBsfermions}
  \Gamma_{\Phi \rightarrow B \tilde{f} \tilde{f}^*} = \frac{ C^2_Y\,m^3_\Phi}
  {32 \pi\Lambda^2}\, \frac {\alpha_Y Y^2_{\tilde{f}}} {32\pi}
  \int \int \frac{ [x^2_3 (1 - 4\mu_{\tilde{f}} - x_3)
    - (1-x_3) (2 - 2x_2 - x_3)^2]} {(1-x_3)^2}\, dx_2 \,dx_3\,,
\end{equation}
with similar integration limits as in (\ref{limits}) but with
$\mu_{\tilde{f}} = m^2_{\tilde{f}}/m^2_\Phi$ instead.

\printbibliography

\end{document}